\documentclass[11pt,a4paper]{article} 
\usepackage{jheppub}

\newcommand{\be}{\begin{equation}}
\newcommand{\ee}{\end{equation}}
\newcommand{\beqa}{\begin{eqnarray}}
\newcommand{\eeqa}{\end{eqnarray}}
\newcommand{\beq}{\begin{eqnarray}}
\newcommand{\eeq}{\end{eqnarray}}
\def\dt{\partial_\tau}

\def\diag{\mathrm{diag}}
\def\dbold{\mbox{\boldmath${\partial}$}}

\newcommand{\lcdm}{$\Lambda$CDM\;}
\newcommand{\lcdmm}{$\Lambda$CDM}
\newcommand{\qa}{Q_{\! A}}
\newcommand{\qm}{Q_{\! M}}

\newcommand{\weff}{\omega_\text{eff}}

\newcommand{\itre}{\text{(i3)}_{\!\phi mr}}
\newcommand{\ifire}{\text{(i4)}_{\!\phi r}}

\newcommand{\iseks}{\text{(i6)}_{\!\phi m}}

\newcommand{\ato}{\text{(a2)}_{\!A\phi r}}
\newcommand{\atre}{\text{(a3)}_{\! A\phi mr}}
\newcommand{\afem}{\text{(a5)}_{\! A\phi m}}

\newcommand{\ito}{\text{RDE(i2)}}
\newcommand{\afire}{\text{A}\phi\text{MDE(a4)}}
\newcommand{\ifem}{\phi\text{MDE(i5)}}
\newcommand{\aseks}{\text{A}\phi\text{DE(a6)}}
\newcommand{\isyv}{\phi\text{DE(i7)}}

\newcommand{\rde}{\text{RDE(i2)}}
\newcommand{\vqmde}{\text{A}\phi\text{MDE(a4)}}
\newcommand{\qmde}{\phi\text{MDE(i5)}}
\newcommand{\vqde}{\text{A}\phi\text{DE(a6)}}
\newcommand{\qde}{\phi\text{DE(i7)}}

\usepackage{graphicx}
\usepackage{bm}
\usepackage{amsmath}
\usepackage{amssymb}

\title{Cosmology of a Scalar Field Coupled  to Matter and  an Isotropy-Violating Maxwell Field }
\author[a]{Mikjel Thorsrud,} \author[a]{David F.~Mota} \author[b]{and Sigbj{\o}rn Hervik}
\affiliation[a]{Institute of Theoretical Astrophysics, University of Oslo, \\ P.O. Box 1029 Blindern, N-0315 Oslo, Norway}
\affiliation[b]{Faculty of Science and Technology, University of Stavanger, \\ Kjell Arholmsgt. 41, N-4036 Stavanger, Norway} 
\emailAdd{mikjel.thorsrud@astro.uio.no}
\emailAdd{d.f.mota@astro.uio.no} 
\emailAdd{sigbjorn.hervik@uis.no}

\abstract{Motivated by the couplings of the dilaton in four-dimensional effective actions, we investigate the cosmological consequences of a scalar field coupled both to matter and a Maxwell-type vector field.  The vector field has a background isotropy-violating component. New anisotropic scaling solutions which can be responsible for the matter and dark energy dominated epochs are identified and explored. For a large parameter region the universe expands almost isotropically. Using that the CMB quadrupole is extremely sensitive to shear, we constrain the ratio of the matter coupling to the vector coupling to be less than $10^{-5}$.  Moreover, we identify a large parameter region, corresponding to a strong vector coupling regime, yielding exciting and viable cosmologies close to the \lcdm limit. }

\begin{document} 
\maketitle

\section{Introduction \label{chintroduction}}
Scalar fields play an important role in novel high energy physics models and in theoretical cosmology. They provide a simple and natural framework for exploring the possibility that the current cosmic acceleration is driven by a dynamical  dark energy component $-$ quintessence - \cite{copeland06}. If not prevented by an unknown symmetry, these fields are expected to couple to other matter fields \cite{carroll98}.  The phenomenology of the coupling to dark matter has been studied extensively in the literature, see for instance \cite{DEboka} and references therein.

A natural possibility that seems to be overlooked in models for the late-time universe, is the coupling of quintessence to vectors. In fact couplings of scalars to Maxwell-type vector fields are quite common in fundamental theories. For instance, there is a scalar-Maxwell coupling in the bosonic sector of the supergravity action. The historically first theory that included both the scalar-matter and scalar-Maxwell couplings was the original Kaluza-Klein model, almost a century ago. Both of these couplings are also present in modern higher dimensional theories such as string theories. It should be stressed that typically both couplings are present simultaneously;  as a concrete example is the low-energy effective gravi-dilaton action \cite{wittenboka} which was explored in the context of quintessence for example in \cite{gasperini01}.  So far, however, the gauge kinetic coupling to the Maxwell field has mostly been ignored in the context of the late-time cosmology.  

The purpose of this work is to explore the cosmological consequences of these couplings for the late-time behaviour of our universe. Although it is natural in the context of higher dimensional theories that both couplings are present simultaneously, it turns out that the cosmology of this doubly-coupled scenario has not been explored in the literature. We shall call the model \emph{Doubly Coupled Quintessence} (DCQ).  Both couplings have, however, separately been studied previously although in very different contexts. There is a long history for studying scalar-matter coupling going back to the early nineties, see \cite{ellis89,damour93,damour94} for early works. We shall refer to the scenario proposed by Amendola in 1999 \cite{amendola} as Standard Coupled Quintessence (SCQ). The main feature of SCQ is a new scaling solution responsible for the matter dominated epoch. The scalar-Maxwell coupling, on the other hand, has not been studied for the late-time cosmology so far, at least not in the context considered here.  In this paper we are primarily interested in the back-reaction on the geometry of the homogenous vector field. In the context of early universe inflation this has been studied extensively recently \cite{Watanabe09,soda10,hervik11,yamamoto12,dimopoulos10,emami10,do2011}. Sourced by the homogenous vector potential, the inflating attractor exhibits a small stable anisotropic hair in the expansion rate which, together with the anisotropic coupling between variables, has interesting phenomenological consequences at the perturbative level \cite{kanno09,watanabe10,gumrukcuoglu10,watanabe10b,emami11,do11,dimopoulos11,soda12,lyth12,dimopoulos12,yamamoto12b}. In this work we shall show that genuinely new behavior arise when both couplings are present simultaneously. The model is phenomenologically very rich and exhibits new scaling solutions that can be responsible for the matter dominated and dark energy dominated epochs. Moreover, we identify a large parameter region that (quantitatively) yields cosmologies close to the \lcdm limit. We put bounds on the model parameters observationally. As we shall see, DCQ can be viewed as a generalization of SCQ and the dynamical trajectories of the universe in the latter model represent special cases of the more general model.

In this paper we shall assume that spatial homogeneity holds. Note that a spatially homogeneous vector field with a non-vanishing background component picks out a preferred direction in the universe and thereby violates isotropy. To consistently study the back reaction of the vector on geometry we shall assume an axisymmetric Bianchi type I metric. This spacetime has homogeneous and flat spatial sections and exhibits a space-like symmetry axis which can be aligned with the vector field (one Hubble expansion rate in the direction parallel to the vector and one in the plane perpendicular to the vector).  It is therefore the simplest spacetime consistent with the symmetries of the matter sector. Clearly, our model violates isotropy, i.e., three-dimensional rotational invariance, one of the pillars of the concordance model of cosmology (\lcdmm).  The cosmic microwave background radiation (CMB) provides evidence that the universe is remarkably close to isotropy. Still one has the interesting possibility for small deviations from the idealized model. In fact, a shear at the one percent level today is consistent with the supernova Ia data \cite{campanelli10}.  As we shall see, however, the CMB provides a far more effective way to constrain the shear in our model, and the upper bound will be significantly lower than what is implied by the supernova Ia data. Since our considered spacetime has the flat Friedmann-Lema\^{i}tre-Robertson-Walker (FLRW) metric as a special case, we can study solutions which are (arbitrarily) close to isotropy and, dynamically, arbitrarily close to \lcdmm. Interestingly, for a large parameter region we shall see that solutions with a small shear are dynamically selected. 

A number of anomalies in observational data hinting towards a violation of rotational invariance have been reported by different groups. There are three famous CMB anomalies reported in the WMAP, namely  the surprisingly low quadrupole \cite{jarosik10} (first seen in COBE), alignment between low multipoles (``axis of evil'') \cite{costa03,schwarz04}, and an hemispherical power asymmetry \cite{eriksen03,eriksen07,hansen08}.  Based on WMAP data a uniform bulk flow of galaxy clusters is reported (``dark flow'') \cite{kashlinsky08,kashlinsky10}.  Other data analyses that indicate a preferred axis include analysis of polarization of electromagnetic radiation propagating over cosmological scales \cite{nodland97,hutsemekers05}, and a ``handedness'' in the orientation of galaxies \cite{longo07}.  It is intriguing that several of the reported anomalies seem to point out a \emph{common} axis in the universe \cite{perivo11,longo07b}. Although the statistical significance is an open issue  still being debated \cite{bennett10, magueijo06,land06}, there is no lack of proposed models (or mechanisms) that violates isotropy at late times \cite{koivisto05,beltran07,koivisto07,koivisto08a,koivisto08,akarsu08,battye09,campanelli11,barrow97, barrow97b, berera03, campanelli06, campanelli07, demianski07, campanelli09,kahniashvili08,longo07b,rodrigues08,koivisto10,barrow85}. The anisotropy is often associated with dark energy and thus imprints in observables usually happens around or after the transition to dark energy domination ($z\lesssim 0.3$).  

The theoretically motivated form of our considered model combined with its rich dynamical behavior clearly distinguishes it from other late-time models. The model is characterized by three dimensionless parameters that controls the strength of the two couplings and the shape of the scalar potential. Leaving the parameters free, we explore the cosmologies that arise in the different parameter regions. For a subspace of parameter space the cosmology is entirely isotropic and dynamically equivalent to SCQ.  In general, however, there will be one or more epochs in the cosmic history where the vector is tracking the background energy density and consequently the universe expands anisotropically. The universe may be anistropic both during the matter and dark energy dominated epochs, or possibly only during one of those epochs. To determine the dynamically preferrable region of parameter space we  use the dynamical system approach and we find that there is a well defined parameter region where the shear is sufficiently small to be consistent with present observations. Interestingly, in this region the background dynamics is close to the \lcdm limit.  Another exotic possibility that occurs in our model (by tuning the model parameters) is that the present state of the universe is the global attractor. This is not possible in SCQ (without introducing time dependent couplings) since it is incompatible with the presence of a matter dominated epoch. In our model this is achieved due to the existence of a new scaling solution which is responsible for the matter dominated epoch.

In principle the homogenous Maxwell field could be identified with a possible uniform component of large-scale cosmic magnetic fields. In our model, however, the coupling to the scalar field would correspond to an extremely rapid time variation in the fine structure constant which would be clearly incompatible with constraints from observations. In this work we do not identify the coupled Maxwell field with the photons of the standard model of particle physics, but assume that it represents a possible sub-dominant dark component of the universe which couples only indirectly to matter fields, i.e., via the scalar field and the metric.  The fact that the present energy density of both dark energy and dark matter exceed the energy density of baryons by roughly one order of magnitude, suggests that we should keep the possibility open for other undiscovered fields with too little energy or too weak couplings to be identified with present data.  

This paper is organized in the following way. In section \ref{chbasics} we introduce the cosmological model and derive the field equations and the autonomous system. Section \ref{chphasespace} is devoted to a general analysis of phase space.  Based on this we identify a very interesting parameter region which we name the \emph{strong vector coupling regime} and explore the cosmology of in section \ref{chstrongvector}.  In section \ref{chconclusion} we summarize and discuss our results. Readers who only want a quick overview of the main results may jump directly to section \ref{chconclusion}. 

\section{Basics of the model \label{chbasics}}

\subsection{Action \label{chaction}}

The doubly coupled model is given by the following Einstein frame action for the metric $g_{\mu\nu}$, scalar field $\phi$, vector field $A_\mu$ and perfect fluids $\Psi_i$:
\begin{equation}
	S = \int d^4x \sqrt{-g} \left[\frac{M_p^2}{2}R(g_{\mu\nu}) + \mathcal L_\phi(g_{\mu\nu}, \phi) + \mathcal L_A(g_{\mu\nu}, A_\mu, \phi) + \mathcal L_M(\tilde g_{\mu\nu}, \Psi_i) \right],
\label{action}
\end{equation}
where $g$ is the determinant of the metric, $M_p$ is the Planck mass and $R$ is the Ricci scalar. The Lagrangians for the scalar field and vector field are, respectively,
\begin{align} 
\mathcal L_\phi &= - \frac{1}{2}g^{\mu\nu} \nabla_\mu \phi \nabla_\nu \phi - V(\phi),\\
\mathcal L_A &= - \frac{1}{4} f^2(\phi)F_{\mu\nu}F^{\mu\nu},
\end{align} 
where the field strength of the vector field is defined by $F_{\mu\nu} = \nabla_\mu A_\nu - \nabla_\nu A_\mu$.   The Jordan frame metric $\tilde g_{\mu\nu}$ is related to the Einstein frame metric by the conformal transformation $\tilde g_{\mu\nu} = h^2(\phi)g_{\mu\nu}$.  Note that the perfect fluid Lagrangian, $\mathcal L_M$, describes matter fields ($\Psi_i$) which are uncoupled in the Jordan frame. In the Einstein frame, on the other hand, these are coupled to the scalar field. Note that apart from the vector part ($\mathcal L_A$), the action is similar to the case of the standard coupled quintessence (SCQ) model \cite{amendola}.  Our doubly coupled model can therefore be viewed as a generalization of SCQ.  
 
Motivated by dimensional reduction of higher dimensional theories such as string theories \cite{martin07} we take the potential of the scalar field and the coupling functions to be exponentials:
\begin{align}
V(\phi)&=V_0 e^{\lambda\frac{\phi}{M_p}},\\
f(\phi)&=f_0 e^{Q_{\! A}\frac{\phi}{M_p}},\\
h(\phi)&=h_0 e^{Q_{\! M} \frac{\phi}{M_p}}, \label{defqm}
\end{align}
where $\lambda$, $\qa$ and $\qm$ are constant parameters characterizing the model.  Once the compactification scheme is specified these can be determined; however, in our model we will leave  them unspecified and explore the cosmologies that arise in the different parameter regions. The simplest (and historically first) model of the form (\ref{action}) is the original Kaluza Klein theory, starting with a cosmological constant in $4+1$ dimensions. The scalar-Maxwell coupling is then a result of the compactification, while the matter coupling arise after a conformal transformation to the Einstein frame.\footnote{In that case the parameters have the values $\lambda\!=\!\sqrt{8}$,$\;\qa\!=\!-\sqrt{6}/2$ and $\qm\!=\!\sqrt{6}/6$. For a discussion of the Kaluza Klein model in the Einstein frame, see \cite{gron07}.}  Note that when referring to SCQ it is understood that we mean Coupled Quintessence with the same exponential type matter coupling and scalar potential as considered here, i.e., the case studied in \cite{amendola}. We consider a positive potential, i.e., $V_0>0$. We set $f_0$ and $h_0$ to unity.\footnote{One can absorb $f_0$ into the scalar field by the rescaling $\phi\rightarrow \phi + \phi_0$. In the limit of vanishing scalar field ($\phi=0$) we demand $g_{\mu\nu}=\tilde g_{\mu\nu}$ implying $h_0=1$.}  

Variations of the action with respect to $g^{\mu\nu}$, $\phi$ and $A_\mu$, lead to the following equations of motion:
\begin{align}
& M_p^2 E^\mu_{\;\nu} = {T}^{\mu(\phi)}_{\;\nu} + {T}^{\mu(A)}_{\;\nu} + {T}^{\mu(M)}_{\;\nu}, \label{eqg} \\
& \square \phi - V'(\phi)  = -Q_A\frac{2\mathcal{L}_A}{M_p} - Q_M \frac{T_M}{M_p}, \label{kgeq} \\
& \nabla_\nu F^{\mu\nu} = -Q_A\frac{2}{M_p}F^{\mu\nu}\nabla_\nu \phi,  \label{Feq}
\end{align}
where $E^\mu_{\;\nu}$ is the Einstein tensor, $T_M=g_{\mu\nu} T^{\mu\nu(M)}$ and $\square \equiv g^{\mu\nu} \nabla_\mu \nabla_\nu$. In the limit $\mathcal{L}_A\rightarrow 0$ the equations agree with the standard formulation of coupled quintessence \cite{amendola}, while in the limit $\mathcal{L}_M\rightarrow 0$ we recover the equations of the inflationary model considered in \cite{soda10,hervik11}. The energy momentum tensor is defined
\be
T^{(i)}_{\mu\nu} = -\frac{2}{\sqrt{-g}} \frac{\delta (\sqrt{-g} \mathcal{L}_{i})}{\delta g^{\mu\nu}},
\ee
and for $\phi$ and $A^\mu$ we get: 
\begin{align}
{T}^{\mu(\phi)}_{\;\nu} &= - \delta^\mu_\nu\left( \frac{1}{2} (\nabla \phi)^2 +V(\phi) \right) + \nabla^\mu \phi \nabla_\nu \phi,   \label{emt1} \\ 
{T}^{\mu(A)}_{\;\nu} &= - f^2(\phi) F^{\mu}_{\;\;\, \alpha}F^{\alpha}_{\;\;\, \nu} - \frac{1}{4}f^2(\phi) \delta^\mu_\nu F_{\alpha\beta}F^{\alpha\beta},  \label{emt2} 
\end{align}
where $(\nabla \phi)^2=g^{\mu\nu}\partial_\mu \phi \partial_\nu \phi$. To keep the matter sector reasonably simple we shall follow most works on coupled quintessence (see \cite{DEboka} and references therein) and use a dust model for the coupled dark matter field to be introduced in the next section. This phenomenological description is a good approximation to any specific model of dark matter that allows a fluid description with negligible pressure, heat-flux and anisotropic stress.  It is evident from the right hand side of the Klein-Gordon equation (\ref{kgeq}) that there is a coupling of the scalar field to matter and to the vector field.  From the action we note that the matter field is uncoupled in the Jordan frame, i.e.:
\be
\tilde \nabla_\mu \tilde T^{\mu(M)}_{\;\nu} = 0.
\label{matterJ}
\ee
Next, we reformulate the matter field equations (\ref{kgeq})-(\ref{Feq}) in terms of the energy momentum tensors, and write the matter equation (\ref{matterJ}) in the Einstein frame:\footnote{ Equations (\ref{eqm1}) and (\ref{eqm2}) are obtained by covariant differentiation of (\ref{emt1}) and (\ref{emt2}) and using the equations of motions (\ref{kgeq}) and (\ref{Feq}).  The identity $\nabla_\gamma F_{\alpha\beta} + \nabla_\alpha F_{\beta\gamma} + \nabla_\beta F_{\gamma\alpha} = 0$ and the relations $\mathcal{L}_M = h^4(\phi) \tilde{\mathcal{L}}_M$ and $T^\mu_{\;\nu} = h^4(\phi) \tilde T^\mu_{\;\nu}$ are useful.}
\begin{align}
\nabla_\mu T^{\mu(\phi)}_{\;\nu} &=  -Q_{\! A} \frac{2\mathcal L_A}{M_p} \nabla_\nu \phi  - Q_{\! M} \frac{T_M}{M_p} \nabla_\nu \phi,  \label{eqm1} \\
\nabla_\mu T^{\mu(A)}_{\;\nu} &= + Q_{\! A} \frac{2\mathcal L_A}{M_p} \nabla_\nu \phi,  \label{eqm2} \\
\nabla_\mu T^{\mu(M)}_{\;\nu} &= + Q_M \frac{T_M}{M_p} \nabla_\nu \phi  \label{eqm3}.
\end{align}
On this form it is clear that the total energy-momentum tensor is covariantly conserved, \newline $\nabla_\mu(T^{\mu(\phi)}_{\;\nu}+T^{\mu(A)}_{\;\nu}+T^{\mu(M)}_{\;\nu})=0$.  It is also manifest that the couplings lead to exchange of energy and momentum both between the scalar field and the vector field, and between the scalar field and the matter field. The rate of exchange is determined by the coupling constants $\qa$ and $\qm$, respectively. Note that matter ``feels'' the vector potential only indirectly via quintessence and the metric.  The equations of motion (\ref{eqm1})-(\ref{eqm3}) give essentially a fluid picture of the doubly coupled dark energy.

\subsection{The cosmological model \label{chmodel}}
In this section we shall introduce the cosmological model to be considered in this paper, and write down the field equations imposing the simplest possible metric compatible with the matter source. In addition to the scalar and vector field, we shall consider the case of two matter fields, pressure-free matter (m) and radiation (r). Thus we have $T^{\mu(M)}_{\;\nu} = T^{\mu(m)}_{\;\nu} + T^{\mu(r)}_{\;\nu}$. As radiation has a tracefree energy momentum tensor, radiation is decoupled from quintessence.  $\qm$ as defined in (\ref{defqm}) therefore represents a universal coupling to all matter fields not being of radiation type. In the context of string theory, however, it is expected that quintessence couples differently to different types of matter fields \cite{gasperini01}.  The usual treatment in the literature is to assume that baryons are uncoupled  while dark matter has one single coupling \cite{DEboka}. We shall neglect baryons in this paper. Since they are both uncoupled and subdominant, this should not change our results at a qualitative level, only slightly quantitatively. Alternatively one can consider $\qm$ as an extremely small universal coupling to both dark matter and baryons (which obviously is a less interesting case since there is no hope of confirming such a small coupling observationally). 

While radiation, pressure-free matter and quintessence are modeled as perfect fluids, the homogeneous vector field violates three-dimensional rotational invariance. Still, the matter sector possesses a rotational symmetry in the space-like plane orthogonal to the vector potential $A_{\mu}$. We shall consider a spatially flat spacetime with the same symmetries as in the matter sector. We choose a coordinate chart such that the symmetry axis is aligned with with the $x$-direction:  
\be
ds^2 = -dt^2 + e^{2\alpha(t)}\left[ e^{-4\sigma(t)}dx^2 + e^{2\sigma(t)}(dy^2+dz^2) \right].
\label{metric}
\ee
Note that the function $e^{\sigma}$ represents the deviation from isotropy, ie. the shear degree of freedom, while $e^{\alpha}$ is the isotropic part. This is a locally rotational symmetric (LRS) Bianchi type I spacetime, the simplest geometry consistent with the matter sector. Our considered geometry and field configuration, ie. a metric possessing a rotational symmetry in the pure space-like plane orthogonal to the vector, is one often assumed in the literature (for an incomplete list, see \cite{boehmer07,Watanabe09,ackerman07,kanno09,soda10,watanabe10,gumrukcuoglu10,dimopoulos10,emami10,do11,do2011}).  In a previous work \cite{hervik11} we discussed the stability and naturalness of this setup. More specifically we considered the most general spatially flat and homogenous universe model, which has one more shear degrees of freedom, and showed that the expansion normalized shear in the space-like plane orthogonal to the vector decays exponentially fast.\footnote{This result holds also in a dust or radiation dominated universe, although the decay is quickest during an inflationary phase. More concretely, it decays as $e^{-(2-q)\alpha}$ where $q$ is the deceleration parameter (\ref{decpar}).} Thus if the assumed LRS symmetry is broken initially, the metric quickly evolves towards an asymptotic limit that agrees with our considered setup.

To quantify the degree of anisotropy in the expansion rate we introduce the dimensionless shear:
\be
\Sigma=\frac{\dot\sigma}{\dot\alpha},
\label{shear}
\ee
where a ``dot'' represents differentiation with respect to cosmic time $t$.  Note that an initial co-moving sphere will evolve into an ellipsoid that can be either  oblate ($\Sigma>0$) or prolate ($\Sigma<0$) depending on the type of matter content.\footnote{The spacetime (\ref{metric}) is sometimes referred to as the ``ellipsoidal universe''. This is, however, a bit misleading since the spatial sections are flat, and not ``ellipsoidal". The ``ellipsoidal universe" can easily be interpreted as the Bianchi type IX model which truely has 3-dimensional ``ellipsoids'' as spatial sections. We will thus refrain from using this term here.} When the shear is sourced by a uniform magnetic or electric field the expansion will be oblate. This is also the case for our coupled Maxwell field.

We shall now introduce a notation that resembles the one used in FLRW cosmology. We let $a$ denote the \emph{geometric mean} of the three scale factors (two of them are equal):
\be
a \equiv e^{\alpha},
\ee
and $H$ the \emph{mean} of the three expansion rates:
\be
H \equiv \dot \alpha.
\ee
Note that  $H=\dot a / a$. We shall refer to $a$ and $H$ as the \emph{mean scale factor} and \emph{mean expansion rate}, respectively (although the former actually is the \emph{geometric} mean). Expressed in terms $a$ and $H$ we can define several quantities on the same form as in the FLRW cosmology.  Note that the proper volume scales proportionally to $a^3$, and we define acceleration as $\ddot a>0$.  We also define the deceleration parameter in the usual way:
\be
q\equiv -1-\frac{\dot H}{H^2},
\label{decpar}
\ee
and the usual implications $q<0 \Leftrightarrow \ddot a>0$ follows.  Next we define a ``redshift'' time parameter on the usual form:
\be
z\equiv -1+\frac{a_0}{a},
\label{isoredshift}
\ee
where $a_0$ is the scale factor today. Note that, since the expansion is anisotropic, the redshift of photons emitted at a constant time slice (such as the last scattering surface) will depend on the direction of the incoming photon. In the appendix we show that, to first order in the anisotropy, $z$ is the redshift averaged over all directions, see (\ref{averagez}) and the following comments. In the cosmological interesting case of a small shear, $z$ can therefore be interpreted as the average redshift. Thus we have generalized the notion of a ``redshift'' time parameter, $z$, to the Bianchi I spacetime and demonstrated a clear physical content. This allows us to refer to a well defined redshift time-parameter.
 
Next, let us consider the matter sector. We assume co-moving fluids and decompose the energy momentum tensor relative to a congruence of fundamental observers with four velocity $u^\mu$ in the standard way \cite{ellis98}:
\be
\begin{split}
& T_{\mu\nu} = (\rho+p)u_\mu u_\nu + pg_{\mu\nu} + s_{\mu\nu}, \\
& s^\mu_{\;\mu} = 0, \quad s_{\mu\nu}=s_{\nu\mu}, \quad s_{\mu\nu}u^\nu =0.
\end{split}
\ee
Here, the energy density $\rho$ and pressure $p$ characterize the perfect part, while the imperfect part is described by the tensor $s_{\mu\nu}$. While matter, radiation and quintessence are perfect fluids, the vector field is imperfect and contributes both to the isotropic pressure $p$ and the anisotropic stress $s_{\mu\nu}$. In this paper we will uniquely work in the coordinate basis defined by the metric (\ref{metric}), and it is understood that the following tensor components are always with respect to this basis.  In the coordinate frame the mixed components for our considered matter sector can be written:
\be
T^\mu_{\;\;\nu} = \text{diag}(-\rho,\; p -2\pi,\; p + \pi, \; p + \pi ),
\label{splitT}
\ee
where $\pi$ represents the anisotropic stress which is related to the energy density of the vector field.  We split the energy density and pressure in the contributions from the scalar field ($\phi$), vector field (A), non-relativistic matter (m) and radiation (r): 
\be
\begin{split}
\rho &= \rho_\phi + \rho_A + \rho_m + \rho_r, \\
p &= p_\phi + p_A + p_m + p_r.
\end{split}
\ee
Then, let us consider each of the matter fields. We will assume the standard matter is pressure-less, i.e., $p_m = 0$, while radiation satisfies the standard equation of state $p_r = \frac{1}{3}\rho_r$. The energy density and pressure of the scalar field takes the standard form 
\be
\rho_\phi = \frac{1}{2}\dot\phi^2 + V(\phi), \quad p_\phi=\frac{1}{2}\dot\phi^2 - V(\phi),  
\ee
with the corresponding equation of state 
\be
\omega_\phi \equiv p_\phi / \rho_\phi = \frac{\dot\phi^2 -2V(\phi)}{\dot \phi^2 + 2V(\phi)}.
\label{wphi}
\ee
For the vector part we shall assume an electric-type field in the $x$ direction (a magnetic type field aligned in the same direction is dynamical equivalent to our considered case). In the gauge $A_0=0$ this corresponds to a vector potential $\mathbf A \equiv A_\mu dx^\mu = A(t) dx$, such that  $\mathbf F \equiv (1/2) F_{\mu\nu} dx^\mu\wedge dx^\nu = \dot A dt \wedge dx$. The energy density, pressure and anisotropic stress of the vector field is:
\begin{equation} 
\rho_A =  \frac{1}{2}\dot A^2 f^2 e^{-2\alpha+4\sigma}, \quad p_A  = \frac{1}{3}\rho_A,  \qquad \pi = \frac{2}{3} \rho_A.
\end{equation}
Also note that, $\mathcal{L}_A = \rho_A$ for our model.

Next, we write down the field equations. The gravitational equation (\ref{eqg})  for our metric (\ref{metric}) yields
\begin{align}
&H^2 = \dot\sigma^2 + \frac{\rho}{3M_p^2} , \label{hconstraint1} \\
&\dot H + 3H^2  =  \frac{1}{2M_p^2}(\rho - p), \label{doth} \\
&\ddot\sigma + 3H\dot\sigma =  \frac{\pi}{M_p^2}, \label{sheareq}
\end{align}
while the matter field equations (\ref{eqm1})-(\ref{eqm3}) become  
\begin{align}
&\dot\rho_\phi + 3H\rho_\phi(1+\omega_\phi) = Q_A \rho_A \frac{2\dot\phi}{M_p} -Q_M\rho_m\frac{\dot\phi}{M_p}, \label{eqmphi} \\
&\dot\rho_A + 4(H+\dot\sigma)\rho_A = -Q_A \rho_A \frac{2\dot\phi}{M_p}, \\
&\dot \rho_m + 3H\rho_m = Q_M\rho_m\frac{\dot\phi}{M_p}, \\
&\dot \rho_r + 4H\rho_r = 0. \label{eqmr}
\end{align}    
From (\ref{sheareq}) it is clear that the shear is sourced by the anisotropic stress of the vector potential.  Also note from (\ref{eqmr}) that radiation is decoupled from $\phi$ since the energy-momentum tensor is traceless, i.e., $g_{\mu\nu}T^{\mu\nu(r)}=0$.  From (\ref{hconstraint1})-(\ref{doth}) it follows that the effective (or total) equation of state is 
\be
\frac{p}{\rho} = -1-\frac{2}{3}\frac{\dot H}{H^2}\frac{1}{1-\Sigma^2} - 2 \frac{\Sigma^2}{1-\Sigma^2}.
\label{poverrho}
\ee
As described in detail in section \ref{chCMB}, constraints from the CMB yield $|\Sigma| \ll 1$; thus  neglecting $\Sigma$ in (\ref{poverrho}) we define the \emph{effective equation of state} parameter:
\be
\weff \equiv -1-\frac{2}{3} \frac{\dot H}{H^2} = \frac{1}{3}(2q-1).
\label{weff}
\ee
For acceptable cosmologies we have $p\simeq\weff \rho$ (note that corrections come first to \emph{second} order in $\Sigma$), and we will define the eras dominated by radiation, matter and dark energy by $\weff \simeq 1/3$, $\weff \simeq 0$ and $\weff \lesssim -1/3$, respectively.  For cosmologies with a large shear, $\weff$ has no significance as an equation of state and is merely another deceleration parameter. It turns out, however, that for a large parameter region, giving viable cosmology, a small shear will be dynamically selected. 

When deriving equations (\ref{hconstraint1})-(\ref{eqmr}) we have neglected the possibility of magnetic type components in $F_{\mu\nu}$. It turns out, however, that if $F_{\mu\nu}$ instead represented a magnetic type field, the only change in the set of equations (\ref{hconstraint1})-(\ref{eqmr}) would be the trivial substitution $\qa\rightarrow -\qa$. A magnetic type field is therefore dynamically equivalent to our considered electric type field.

\subsection{Autonomous system \label{chaut}}
To study the phase space structure we now introduce dimensionless variables and write down the equations of motion as an autonomous system of equations. The density parameters are defined
\be
\Omega_i \equiv \frac{\rho_i}{3H^2M_p^2},
\ee
where the $i$ denote the vector field (A), scalar field ($\phi$), matter field ($m$) or radiation ($r$).  The Hamiltonian constraint equation (\ref{hconstraint1}) can then be written on the generic form
\be
1=\Sigma^2 + \Omega_A + \Omega_\phi + \Omega_m + \Omega_r,
\label{hconstraint2}
\ee 
where $\Sigma$ is the shear degree of freedom  (\ref{shear}). It is useful to split the scalar in the kinetic energy and potential:
\be
\Omega_\phi = \frac{1}{6} X^2 + \Omega_V,
\label{hconstraint22}
\ee
where 
\be
\quad X \equiv \frac{\dot \phi}{HM_p}
\label{X}
\ee
and
\be
\Omega_V \equiv \frac{V(\phi)}{3H^2M_p^2}. 
\ee
We will use the constraint (\ref{hconstraint2}) to eliminate $V(\phi)$ from the equations of motion.  The system can then be written as a autonomous set of first order differential equations in terms of the \emph{independent variables}: $X$, $\Sigma$, $\Omega_A$, $\Omega_m$ and $\Omega_r$.  It will prove useful to express the equation of state parameters, defined in (\ref{wphi}) and (\ref{weff}), in terms of these variables:
\be
\omega_\phi = \frac{X^2-6\Omega_V}{X^2+6\Omega_V},
\ee
\be
\omega_\text{eff} = -1+ \frac{1}{3}X^2 + 2\Sigma^2 + \frac{4}{3}\Omega_A + \Omega_m + \frac{4}{3} \Omega_r,
\ee
where in the former it is understood that $\Omega_V$ is a function of the independent variables.
  
We now switch to the dimension less time variable $\alpha$ by use of the identity $\frac{d\alpha}{dt}=H$. The autonomous equations can then be written:
\begin{align} 
 \frac{dX}{d\alpha} &= (X+\lambda)\left(  3(\Sigma^2-1) + \frac{1}{2}X^2  \right) + 2X\Omega_A + 3(2Q_A+\lambda)\Omega_A \label{seqX}\\ 
&\qquad+3\lambda(\Omega_m+\Omega_r) +\frac{3}{2}\Omega_m X + 2\Omega_r X -3Q_M\Omega_m,\notag \\
\frac{d\Sigma}{d\alpha} &= 2\Omega_A(\Sigma+1) + \Sigma \left[ 3(\Sigma^2-1) + \frac{1}{2} X^2 +\frac{3}{2}\Omega_m +2\Omega_r \right], \label{seqSigma}\\
\frac{d\Omega_A}{d\alpha} &= 2\Omega_A \left[  3(\Sigma^2-1) + \frac{1}{2}X^2 - Q_A X + 1 - 2\Sigma + 2\Omega_A +\frac{3}{2}\Omega_m + 2\Omega_r \right], \label{seqA}\\
\frac{d\Omega_m}{d\alpha} &= \Omega_m \left[-3 + 6\Sigma^2 + X^2 + Q_M X + 4 \Omega_A + 3\Omega_m + 4\Omega_r)  \right], \label{seqm}\\
\frac{d\Omega_r}{d\alpha} &= \Omega_r \left[-4 + 6\Sigma^2 + X^2 + 4 \Omega_A + 3\Omega_m + 4\Omega_r)  \right].  \label{seqr} 
\end{align}
Since we are considering a positive potential ($V(\phi)\ge0\Rightarrow \Omega_V\ge 0$ ), it follows from (\ref{hconstraint2})- (\ref{hconstraint22}) that the dynamical variables are subject to the constraint
\be
\frac{1}{6}X^2 + \Sigma^2 + \Omega_A + \Omega_m + \Omega_r \le 1.
\label{constraint}
\ee
Since $\Omega_i\ge0$ for all $i$ this implies an upper bound on each variable individually. We shall often write $\Omega_V\ge 0$ instead of referring to (\ref{constraint}). Finally we also write down the equation of motion for the auxiliary variable, $\Omega_V$, which is also useful:\footnote{In simulations, for numerical reasons, it is sometimes useful to replace one of the variables in the autonomous system by $\Omega_V$.  In particular this is the case if the initial condition for $\Omega_V$ is (very) small. Equation (\ref{seqV}) can also be used as a consistency check that the autonomous system (\ref{seqX})-(\ref{seqr}) is correct.}
\be
\frac{d\Omega_V}{d\alpha} = \Omega_V \left( \lambda X +3 + 3 \weff \right) = \Omega_V \left( \lambda X + X^2 + 6\Sigma^2 + 4\Omega_A + 3\Omega_m +4\Omega_r \right).
\label{seqV}
\ee

\section{Phasespace analysis \label{chphasespace}}

Equipped with the autonomous system we shall now investigate the phase-space structure using a dynamical system approach. In section \ref{DSoverview} we identify the fix-points of the system and discuss some of their main properties.  In \ref{DSisotropic} and \ref{DSanisotropic} we study the isotropic and anisotropic solutions, respectively, in greater detail; investigate their physical properties, classify their stabilities and determine the conditions for existence. The purpose is to identify, or rule out, new interesting cosmological scenarios. In \ref{DSattractors}, we characterize parameter space in terms of the attractors.  Finally, in \ref{chgeneric}, we investigate the generality of the stability of the inflationary background solutions.

As we shall see, the phase space of our model is very rich with 7 isotropic and 6 anisotropic fix-points which are all physically different. Since the properties of these solutions depend on a three dimensional parameter space ($\qm$,$\qa$,$\lambda$), a mathematically complete investigation would include extremely complicated algebraic expressions which are not very informative. In this paper we are mostly interested in the cosmology, and therefore, whenever it is desirable and possible, we shall use bounds from observations to simplify the discussion. We will thus not emphasize mathematical completeness, and will avoid  describing the cosmology of models having parameters clearly incompatible with observations, for instance the case of a strong matter coupling, $|\qm|\gg 1$. Using first order approximations will enable us to rewrite most expressions on a simple and informative form applicable to extract cosmological results.

\subsection{Overview \label{DSoverview}}

The fix-points of the system are found by setting the left-hand side of the autonomous system (\ref{seqX})-(\ref{seqr}) equal to zero and solving the algebraic equations. The stability is determined, except in degenerate cases, by linearizing the field equations around the fix-points, $\frac{d \delta X^i}{d\alpha}=\mathcal{M}\delta X^i$, and evaluating the eigenvalues of the matrix $\mathcal{M}$. 

\begin{table}
\centering
\footnotesize
\begin{tabular}{l*{8}{c}}
\hline\hline
         & (i1$\pm$)   & $\ito$  & (i3) & (i4) & $\ifem$& (i6) & $\isyv$  \\ \hline
$X$  & $\pm\sqrt{6}$ & $0$ & $-\frac{1}{\qm}$ & $-\frac{4}{\lambda}$ &  $-2\qm$ & $\frac{3}{\qm-\lambda}$ & $-\lambda$   \\
$\Sigma$   & $0$ & $0$ & $0$ & $0$ &  $0$ & $0$ & $0$   \\
$\Omega_A$ & $0$ & $0$ & $0$ & $0$ &  $0$ & $0$ & $0$   \\
$\Omega_m$ & $0$ & $0$ & $\frac{1}{3\qm^2}$ & $0$ &  $1-\frac{2\qm^2}{3}$ & $\frac{-3-\lambda\qm+\lambda^2}{(\qm-\lambda)^2}$ & $0$   \\
$\Omega_r$ & $0$ & $1$ & $1-\frac{1}{2\qm^2}$ & $1-\frac{4}{\lambda^2}$ &  $0$ & $0$ & $0$   \\
$\Omega_\phi$ & $1$ & $0$ & $\frac{1}{6\qm^2}$ & $\frac{4}{\lambda^2}$ &  $\frac{2\qm^2}{3}$ & $\frac{\qm(\qm-\lambda)+3}{(\qm-\lambda)^2}$ & $1$   \\
$\omega_\phi$ & $1$ & $-$ & $1$ & $\frac{1}{3}$ &  $1$ & $\frac{-\qm(\qm-\lambda)}{\qm(\qm-\lambda)+3}$ & $-1+\frac{\lambda^2}{3}$   \\
$\omega_\text{eff}$ & $1$ & $\frac{1}{3}$ & $\frac{1}{3}$ & $\frac{1}{3}$ &  $\frac{2\qm^2}{3}$ & $\frac{-\qm}{\qm-\lambda}$ & $-1+\frac{\lambda^2}{3}$   \\
\hline\hline
\end{tabular}
\caption{The isotropic fix-points.}
\label{tabis}
\end{table}
\begin{table}
\centering
\footnotesize
\begin{tabular}{l*{8}{c}}
\hline\hline
         & (a1$\pm$) & (a2) & (a3) & $\afire$    \\
\hline
$X$  & $\text{free}$ & $-\frac{4}{\lambda}$ & $-\frac{1}{\qm}$ & $-\frac{3(\qa+3\qm)}{4+(2\qa+\qm)(3\qa+\qm)}$  \\
$\Sigma$  & $\pm\sqrt{1-\frac{X^2}{6}}$ & $\frac{2\qa}{\lambda}$ & $\frac{\qa}{2\qm}$ & $\frac{-1+2\qm(2\qa+\qm)}{4+(2\qa+\qm)(3\qa+\qm)}$    \\
$\Omega_A$ & $0$ & $\frac{\qa}{\lambda}$ & $\frac{\qa}{4\qm}$ & $\frac{3}{2}\frac{\left[2+(3\qa-\qm)(\qa+\qm)\right]\left[-1+2\qm(2\qa+\qm)\right] }{\left[4+(2\qa+\qm)(3\qa+\qm)\right]^2}$   \\
$\Omega_m$ & $0$ & $0$ & $\frac{2+3\qa^2}{6\qm^2}$ & $3\frac{\left[2+(3\qa-\qm)(\qa+\qm)\right]\left[3+2\qa(2\qa+\qm)\right]}{\left[4+(2\qa+\qm)(3\qa+\qm)\right]^2}$   \\
$\Omega_r$ & $0$ & $\frac{\lambda^2-\lambda\qa-6\qa^2-4}{\lambda^2}$ & $\frac{4\qm^2 -\qm\qa -3\qa^2 -2}{4\qm^2}$ & $0$   \\
$\Omega_\phi$ & $1-\Sigma^2$ & $2\frac{2+\qa^2}{\lambda^2}$ & $\frac{1}{6\qm^2}$ & $\frac{3}{2}\frac{(\qa+3\qm)^2}{\left[ 4+(2\qa+\qm)(3\qa+\qm) \right]^2}$   \\
$\omega_\phi$ & $1$ & $-1+\frac{8}{3(2+\qa^2)}$ & $1$ & $1$  \\
$\omega_\text{eff}$ & $1$ & $\frac{1}{3}$ & $\frac{1}{3}$ & $\frac{\qm(\qa+3\qm)}{4+(2\qa+\qm)(3\qa+\qm)}$   \\
\hline\hline
\end{tabular}

\begin{tabular}{l*{4}{c}}
         & ($a5$)   & $\aseks$    \\
\hline
$X$ & $\frac{3}{\qm-\lambda}$  & $-\frac{12(2\qa+\lambda)}{8+(2\qa+\lambda)(6\qa+\lambda)}$     \\
$\Sigma$ & $\frac{\lambda-6\qa-4\qm}{4(\qm-\lambda)}$  & $\frac{2(\lambda^2+2\lambda\qa-4)}{8+(2\qa+\lambda)(6\qa+\lambda)}$    \\
$\Omega_A$ & $-3\frac{(2\qm-\lambda)(6\qa+4\qm-\lambda)}{16(\qm-\lambda)^2}$ & $3\frac{\left[\lambda^2+2\lambda\qa-4\right]\left[8+(6\qa-\lambda)(2\qa+\lambda)\right]}{(8+(2\qa+\lambda)(6\qa+\lambda))^2}$   \\
$\Omega_m$ & $-3\frac{8+(6\qa+4\qm-3\lambda)(2\qa+\lambda)}{8(\qm-\lambda)^2}$ & $0$   \\
$\Omega_r$ & $0$ & $0$  \\
$\Omega_\phi$ & $\frac{3}{8}\frac{8+6\qa^2+6\qa\qm+4\qm^2-\lambda(\qa+3\qm)}{(\qm-\lambda)^2}$ & $6\frac{16 + 56 \qa^2 + 24 \qa^4 + 32 \qa \lambda + 20 \qa^3 \lambda + 
 2 \lambda^2 + 2 \qa^2 \lambda^2 - \qa \lambda^3}{(8+(2\qa+\lambda)(6\qa+\lambda))^2}$  \\
$\omega_\phi$ & $-1+\frac{8}{8+6\qa^2+6\qa\qm+4\qm^2-\lambda(\qa+3\qm)}$ & $-1+\frac{8 (2 \qa + \lambda)^2}{16 + 56 \qa^2 + 24 \qa^4 + 32 \qa \lambda + 20 \qa^3 \lambda + 
 2 \lambda^2 + 2 \qa^2 \lambda^2 - \qa \lambda^3}$   \\
$\omega_\text{eff}$ & $-\frac{\qm}{\qm-\lambda}$ & $-\frac{8+12\qa^2-3\lambda^2}{8+(2\qa+\lambda)(6\qa+\lambda)}$   \\
\hline\hline
\end{tabular}
\caption{The anisotropic fix-points.}
\label{taban}
\end{table}

We find $7$ isotropic fix-points (i1)-(i7), and $6$ anisotropic fix-points (a1)-(a6), see tables \ref{tabis} and \ref{taban} respectively. In these tables the independent variables ($X$, $\Sigma$, $\Omega_A$, $\Omega_m$, $\Omega_r$) are specified, together with $\omega_\phi$, $\omega_\text{eff}$ and the auxiliary variable $\Omega_\phi$. Especially $\omega_\text{eff}$ characterize much of the physics of these solutions. A radiation dominated solution must have $\weff\simeq 1/3$, a matter dominated solution $\weff\simeq 0$, while $\weff < -1/3$ for an accelerated (dark energy dominated) solution.  In each of the tables the fix-points are sorted roughly in terms if decreasing $\weff$. The isotropic fix-points (i1)-(i7) also exist in, and completely characterize, the standard coupled quintessence (SCQ).\footnote{When we compare our model to CQ it is understood that we mean CQ with the same type of scalar potential and matter coupling as in our model, i.e., exponential functions linear in $\phi$. The relation between our notation and the parameters $\beta$ and $\mu$ in \cite{amendola} are $\beta=-\frac{\sqrt{6}}{2}\qm$ and $\mu=\frac{\sqrt{6}}{2}\lambda$, while the relation to the notation in \cite{DEboka} is simply $\lambda\rightarrow -\lambda$.}  These well known solutions are carefully analyzed elsewhere \cite{amendola,DEboka}, but since we have two more degrees of freedom ($\Sigma$ and $\Omega_A$), the stability will in general change and needs to be reanalyzed. The phase-space of our model is richer than SCQ since we also have the anisotropic fix-points (a1)-(a6). The scaling solutions (a2), (a3), (a4) and (a5) are genuinely new, while (a6) also exists in the inflation model studied in \cite{dimopoulos10,soda10,hervik11}. Apart from (a1) (which is of little relevance), non of them have previously been considered in the context of late-time cosmology. 

In order to have a self explanatory notation we shall often label the fix-points by $A$, $\phi$, $m$ and/or $r$ to show which fluids they contain. For instance we shall often write $\atre$ to remind the reader that (a4) contains vector (A), scalar ($\phi$), matter (m) and radiation (r).  Note that $\Sigma=\Omega_A=0$ for \emph{all} isotropic fix-points (i1)-(i7).  Five of the fix-points turns out to be so important that we have introduced acronyms beyond this labeling. This is $\ito$, $\afire$, $\ifem$,  $\aseks$ and $\isyv$.  The two last letters in the acronyms denotes ``Dominated Epoch''; for instance ``Radiation Dominated Epoch ($\ito$)'' and ``Vector-$\phi$-Matter Dominated Epoch ($\afire$)''.  This type of acronym was introduced by Amendola (in \cite{amendola}) for the fix-point $\ifem$ which represents the main feature of SCQ.  We find it natural to introduce the same sort of acronyms in this generalized model.

 In the following two sections we shall determine the stability and existence of the fix-points. In general the stability and existence depend on the constant parameters $\lambda$, $\qa$ and $\qm$. Our goal is to get insight into the model by considering different regimes for the parameters, and identify those that lead to interesting and viable cosmologies.  A given fix-point exists in the part of parameter space satisfying the following conditions: $X$ and $\Sigma$ must be real,  ($\Omega_A$, $\Omega_m$, $\Omega_r$) must be non-negative \emph{and} the inequality (\ref{constraint}) must hold.\footnote{These conditions also imply that $\Omega_\phi\ge 0$.} This gives in principle six different conditions for existence, but in all cases  some of them are automatically satisfied while others are related. The independent number of conditions on ($\lambda$, $\qm$, $\qa$) is therefore reduced to maximum three.   Without loss of generality we consider $\lambda>0$ in the following analysis.\footnote{Negative $\lambda$ is equivalent to the situation $\phi \rightarrow -\phi$.}

\paragraph{Exact solutions} The fix-points correspond to exact power law solutions which are straight forward to derive. From the matter field equations (\ref{eqmphi})-(\ref{eqmr}) we get:
\be
\rho_i \propto a^{-\beta_i},
\label{eksakt1}
\ee
where
\begin{equation} 
\beta_i =
\begin{cases} 
3+3\omega_\phi -2\qa X \frac{\Omega_A}{\Omega_\phi} + \qm X \frac{\Omega_m}{\Omega_\phi}, & \text{for }i= \phi,  \\
4+4\Sigma + 2\qa X, & \text{for }i= A, \\
3-\qm X, & \text{for }i= m, \\ 
4, & \text{for }i= r. \\ 
\end{cases} 
\label{eksakt2}
\end{equation}
The fact that all fluids scales similarly in the fix-points, implies that $\beta_i(\qm,\qa,\lambda) = \beta_j(\qm,\qa,\lambda)$  where $i$ and $j$ represents two arbitrary fluids of the considered fix-point.\footnote{To avoid possible confusion let us elaborate a bit here. For a fix-point that contains radiation, $\beta_i=4$ for all fluids contained in that fix-point. For fix-points \emph{not} containing radiation, $\beta_i$ will be parameter dependent and in general of course different from $4$. The point is that all fluids that exist in a given fix-point scale similarly and thus the \emph{corresponding} coefficients $\beta_i$ are equal.} We therefore write $\beta_i=\beta$. It is then straight forward to integrate up (\ref{hconstraint1}) and solve for the mean scale factor:
\be
a \propto t^{2/ \beta}.
\label{eksakt3}
\ee

\subsection{Isotropic solutions\label{DSisotropic}}
Here we shall discuss the isotropic fix-points given in table \ref{tabis}. As mentioned above these fix-points completely characterize SCQ, and we will mention which labels/names they are given in \cite{amendola} and \cite{DEboka}.  Note that the stability will in general be different from SCQ since we have two more degrees of freedom ($\Sigma$ and $\Omega_A$). 

\paragraph{Fixpoint (i1$\pm)_\phi$}  This fix-point contains only a pure kinetic scalar field. Thus the fluid is ``stiff'' ($\weff=1$), and therefore dynamically irrelevant (since there is no ``stiff'' epoch in the cosmic history). (i1$+$) and (i1$-$), respectively, are labeled (e) and (d) in \cite{amendola}, and (b1) and (b2) in \cite{DEboka}. Like in SCQ the fix-point is unstable, and exist in the entire parameter space.


\paragraph{Fixpoint $\rde$} This fix-point contains only pure radiation ($\Omega_r=1$) and corresponds to the radiation dominated epoch both in \lcdm and SCQ.  We shall refer to it as the ``Radiation Dominated Epoch ($\ito$)'', and it corresponds to fix-point ($c_R$) in \cite{amendola} and (e) in \cite{DEboka}. As we shall see in section \ref{chinitialconditions}, cosmologically viable initial conditions must start close to this fix-point also in our model. It is (obviously) a saddle and exists in the entire parameter space ($\lambda,\qm,\qa$).


\paragraph{Fixpoint $\itre$} This fix-point represents a scaling solution containing a pure kinetic scalar, matter and radiation. It is labeled ($c_{RM}$) in \cite{amendola} and (f) in \cite{DEboka}. It exists in the parameter region $\qm^2 \ge 1/2$. Apparently, with $\weff=1/3$  it is a candidate for the radiation dominated epoch.  As we shall see in section \ref{chinitialconditions}, however, initial conditions starting close to this fix-point will not realize a sufficiently stable radiation dominated epoch  lasting (as a minimum) from big-bang nucleosynthesis (BBN) all the way to redshift $z\sim 3000$. This rules out (i3) as a candidate for the radiation dominated epoch.\footnote{Here we should mention that, although for a different reason, this fix-point is also ruled out as a candidate for the radiation dominated epoch in the framework of SCQ \cite{DEboka}. In that case the problem is the BBN bounds on primordial quintessence ($\Omega_\phi^\text{bbn}<0.045$ \cite{bean01}) which implies $\qm^2>3.7$. Such a large matter coupling is incompatible with the existence of a subsequent matter dominated epoch (which must be provided by $\ifem$ in SCQ). Note that this argument does not hold in the case of our more general model since it is possible to realize a matter dominated epoch with such a large matter coupling via the new scaling solution $\afire$.}

Let us also comment on the stability. According to the eigenvalues ($\ref{eigeni3}$), (i3) is stable in the intersection of $\frac{\qa}{\qm} < 0$, $\frac{\lambda}{\qm}>4$ and $\qm^2>\frac{1}{2}$. Elsewhere it is a saddle. Also in SCQ the stability depends on the parameters and (i3) will either be an attractor or a saddle.


\paragraph{Fixpoint $\ifire$} This fix-point represents a scaling solution with quintessence ($\omega_\phi=1/3$) and radiation. It is labeled ($b_R$) in \cite{amendola} and (g) in \cite{DEboka}.  It exists in the parameter region $\lambda \ge 2$ and is stable in the intersection of $\qa<0$ and $4\qm>\lambda$.\footnote{Note that the real part of the two last eigenvalues in (\ref{eigeni4}) are always negative in the parameter region where the fix-point exists.} Like (i3) it has $\weff=1/3$ and is also apparently a candidate for the radiation dominated epoch.  As shown in section \ref{chinitialconditions}, however, initial conditions starting close to this fix-point cannot provide a sufficiently stable matter dominated epoch lasting all the way from $z\sim 3000$ up to transition to dark energy domination at low redshift $z\sim 1$. Therefore also (i4) is ruled out as a candidate for the radiation dominated epoch.\footnote{Although for a different reason, also in the framework of SCQ this fix-point is ruled out as candidate for the radiation dominated epoch \cite{DEboka}. In that case the problem is that the BBN bounds on a primordial quintessence (mentioned in the above paragraph) implies $\lambda^2 > 88.9$ which is incompatible with a dark energy dominated epoch (which must be realized by $\isyv$). Note that this argument does not hold in the framework of our more general model since a viable dark energy dominated epoch can be realized by the scaling solution $\aseks$ as long as the the fraction $|\lambda/\qa|$ is small.} 


\paragraph{Fixpoint $\qmde$} This is the well known scaling solution responsible for the matter dominated epoch in SCQ. It is quite routinely referred to as the ``$\phi$-Matter Dominated Epoch ($\phi$MDE)'', and is labeled ($c_M$) in \cite{amendola} and (a) in \cite{DEboka}.  It exists in the parameter region $\qm^2 \le 3/2$. However, $|\qm| < 0.13$ (which corresponds to $\Omega_\phi<0.011$) is required to satisfy CMB constraints \cite{amendola03}.  The phenomenology of this fix-point is well understood at the (linear) perturbative level.  For instance matter fluctuations grows less than in the uncoupled case, CMB acoustic peaks are shifted to higher multipoles compared to \lcdm (due to change in the sound horizon) and low multipoles are tilted (due to the integrated Sachs-Wolfe effect) \cite{amendola}. Regarding the stability, it is interesting to note that that the fourth eigenvalue in (\ref{eigeni5}) is positive in the parameter region where the fix-point $\afire$ exists. This reflects the fact that in the parameter regions where both $\ifem$ and $\afire$ exist, the latter one will be dynamically selected.  In section \ref{chdynamics} we shall see that in general $\afire$ will be responsible for the matter dominated epoch, while $\ifem$ will be responsible only in the parameter region where $\afire$ does not exist.  For $|\qm|<0.13$ $\ifem$ will in general be a saddle followed by a late time accelerated solution. However, in the parameter region where $\afire$ does not exist, it will be a stable attractor provided the additional requirements $\qm>0$ and $\lambda>\frac{3+2\qm^2}{2\qm}$. Obviously, this parameter region (notice that $\lambda\gg 0$ when $\qm\ll1$) can be ruled out since it will be impossible to realize the late-time accelerated epoch.

\paragraph{Fixpoint $\iseks$} This is a scaling solution with matter and quintessence. It is labeled ($b_M$) in \cite{amendola} and (d) in \cite{DEboka}. In principle (i6) is a candidate for both the matter and dark energy dominated epoch. In the framework of SCQ, (i6) can be ruled as a viable candidate for any of these periods \cite{DEboka,amendola}. Below we show that in our more general model the same conclusion holds only for the matter dominated epoch.  Quite interestingly, it is possible to realize a viable cosmology where (i6) represents an accelerated late-time attractor.  The fix-point exists in the parameter region where $2\qm(\lambda-\qm) \le 3$ and $\lambda(\lambda-\qm)\ge 3$, where the first condition comes from $\Omega_V>0$ and the latter from $\Omega_m>0$.   

Then, let us show why (i6) is not a viable candidate for the matter dominated epoch.  In the limit $\lambda \gg |\qm|$ and $\lambda\gg1$ the fix-point is apparently a candidate for the matter dominated epoch with $\weff\simeq 0$ and $\Omega_m \simeq 1$. In order to have a subsequent accelerated epoch we must further require $6\qa \lesssim \lambda$ so that we have saddle stability (follows from a detailed study of the eigenvalues (\ref{eigeni6}) combined with the existence conditions).  Since $\lambda\gg1$, the only possibility for a subsequent accelerated epoch is provided by $\aseks$.  Then, $|\qa|\gg \lambda$ is required in order to have a small shear in the accelerated era. Combined with the above conditions we then have $-\qa \gg \lambda \gg 1$. In that case $\aseks$ does not exist.  Thus it is impossible to realize a subsequent accelerated epoch. This rules out $\iseks$ as a possible candidate for the matter dominated epoch.    

Next, let us consider (a6) as a candidate for the accelerated epoch. Acceleration  ($\weff \in [-1,-\frac{1}{3})$) is realized in the parameter region $\qm<-\frac{1}{2}\lambda$.\footnote{For $\qm>\lambda$ we have $\weff<-1$ and the fix-point does not exist.}  Under this condition the two first eigenvalues in (\ref{eigeni6}) is negative. If we additionally have $\qa<\frac{\lambda-4\qm}{6}$ and $\qm<\lambda-\frac{3}{\lambda}$, also the other eigenvalues are negative. In that parameter region $\iseks$ represents an accelerated scaling solution with matter and quintessence.  This includes the special case $(\qm,\,\lambda)=(-3.3,\, 2.2)$ in which case $\Omega_\phi \simeq 0.7$ and $\Omega_m \simeq 0.3$. This opens up the exotic possibility that the present state of the universe represents a global attractor. Interestingly, in this scenario the universe has already reached the final attractor solution and is not in the middle of the transition between two cosmic epochs. As mentioned above, in the framework of SCQ this possibility is inconsistent with the presence of a matter dominated epoch since $\ifem$ does not exist in the considered point in parameter space \cite{amendola,DEboka}. In our more general model, however, the matter dominated epoch can instead be realized by the scaling solution $\afire$, and, quite remarkably, it is indeed possible to realize a viable cosmology on the background level.  We consider the details of this possibility in section \ref{chpresent}.

\paragraph{Fixpoint $\isyv$}  This is the well known quintessence attractor which exists when $\lambda^2 \le 6$ and accelerates when $\lambda^2 < 2$. It is labeled (a) in \cite{amendola} and (c) in \cite{DEboka}. Note that it represents the late-time accelerated solution both in SCQ and in ``conventional'' quintessence without matter coupling (with exponential type potential). In our more general model it is one of three possible late-time attractors that can realize acceleration  (together with $\iseks$ and $\aseks$).  From the eigenvalues (\ref{eigeni7}) it follows that the fix-point is stable and an attractor in the intersection of $\lambda^2<4$, $\qm>\lambda-\frac{3}{\lambda}$ and $\qa<\frac{4-\lambda^2}{2\lambda}$.


\subsection{Anisotropic solutions \label{DSanisotropic}}
Here we shall discuss the anisotropic fix-points given in table \ref{taban}. As mentioned above, (a1) is a well-known solution of little relevance; (a2),(a3),(a4) and (a5) are genuinely new scaling solutions; while (a6) is explored in the context of inflation \cite{dimopoulos10,soda10,hervik11}. Apart from (a1), non of them are considered in the context of late-time cosmology.    

\paragraph{Fixpoint (a1$\pm)_\phi$}
This is a decelerating fix-point containing only a pure kinetic scalar field (thus the fluid is ``stiff'' $\rho=p$). Since there is no ``stiff'' epoch in the cosmic history, this solution is dynamically irrelevant. The solution is part of a broader solution commonly referred to as \emph{Jacobs disc} \cite{jacobs68}.  See also \cite{hervik01} for a discussion of such solutions.  Note that (e) is a \textit{curve} of fix-points, satisfying $\frac{1}{6}X^2 + \Sigma^2=1$. Notice from the eigenvalues (\ref{eigena1}) that the fix-point is unstable.

\paragraph{Fixpoint $\ato$}  
This is a genuinely new scaling solution containing radiation, vector and quintessence. With $\weff = 1/3$ it is apparently a candidate for the radiation dominated epoch. However, in (\ref{chinitialconditions}) we shall see that it is impossible to realize a sufficiently stable radiation dominated epoch lasting (as a minimum) from big-bang nucleosynthesis ($T_{\text{bbn}} \sim 1 \text{MeV}$) all the way to redshift $z\sim 3000$. 

The fix-point exists in the parameter region where $\qa\ge 0$ and $\lambda\ge \frac{1}{2}\qa +\frac{1}{2}\sqrt{25\qa^2+16}$. The former follows from $\Omega_A\ge 0$, while the latter from $\Omega_r>0$. Notice that the fix-point does not exist in the parameter region $\lambda<2$. The fix-point is a saddle in the major fraction of the parameter space ($\qm, \qa, \lambda$), but note that there are certain spots where it is a stable attractor, see eigenvalues (\ref{eigena2}).

\paragraph{Fixpoint $\atre$}  This is a scaling solution with radiation, matter, vector and the kinetic part of the scalar. With $\weff=1/3$ it is apperantly a candidate for the radiation dominated era. However, in section \ref{chinitialconditions} we show that it is impossible to realize a subsequent viable matter dominated epoch.  Thus we can rule out (a3) as a candidate for the radiation era.  

The fix-point exists in the intersection of $\frac{\qa}{\qm} > 0$ and $4\qm^2-\qm\qa-3\qa^2>2$. The former follows from $\Omega_A>0$, while the latter from $\Omega_r>0$.  Thus the fix-point does not exist for a small matter coupling $|\qm|\ll1$, more precisely it does not exist in the region $\qm\in (-\frac{\sqrt{2}}{2},\frac{\sqrt{2}}{2})$. Like (a2) this fix-point is in general a saddle, but in certain spots of parameter space ($\qm, \qa, \lambda$) it is a stable attractor, see eigenvalues (\ref{eigena3}).

\paragraph{Fixpoint $\vqmde$} This is a genuinely new scaling solution with matter, vector and the kinetic part of the scalar field.  As we shall show below this fix-point is a candidate for the matter dominated epoch in the limit of a strong vector coupling.  In chapter \ref{chdynamics} we shall see that, whenever it exists, $\vqmde$ will be the dynamically selected solution during the matter dominated epoch. We shall refer to (a4) as the ``Vector-$\phi$-Matter Dominated Epoch ($\afire$)''.  The exact eigenvalues are given in (\ref{eigena4}). In the analysis below we shall focus on the parameter region where $\vqmde$ represents a matter dominated solution.


The conditions for existence follows from $\Omega_A>0$ and $\Omega_m>0$.  For  $|\qm|<\sqrt{6}/2$ this gives the single condition $4\qm\qa +2\qm^2>1$.  For a small matter coupling ($|\qm|\ll1$) this condition reads $|\qa|\gg1$ (with the sign of $\qa$ equal to the sign of $\qm$). In this region we have to lowest order $\Omega_m=1$ and $\Omega_\text{eff}=0$, which means that the fix-point is a candidate for the matter-dominated epoch.  Under the assumption $|\qa| \gg |\qm|$ and $|\qa|\gg1$ we expand the characteristics of the fix-point to first order in the small quantities $\frac{1}{\qa^2}$ and $\frac{\qm}{\qa}$: 
\be
\begin{split}
&X=-\frac{1}{2\qa}, \quad \Sigma=\frac{2\qm}{3\qa}-\frac{1}{6\qa^2}, \quad \Omega_A=\frac{\qm}{2\qa}-\frac{1}{8\qa^2},\quad \\ & \Omega_m=1-\frac{\qm}{2\qa} +\frac{1}{12\qa^2}, \quad \Omega_r=0, \quad \Omega_\phi =\frac{1}{24\qa^2}, \quad \omega_\phi=1 , \quad \omega_\text{eff} = \frac{\qm}{6\qa}.
\end{split}
\label{a4quantities}
\ee
Thus, to zeroth order, $\vqmde$ looks like a standard matter dominated epoch with $\Omega_m=1$. In section \ref{chCMB} we shall constraint the ratio $\qm/\qa$ observationally by requiring a small $\Sigma$ compatible with observations. Applying equations (\ref{eksakt1})-(\ref{eksakt3}) we find, in the same approximation, that the energy density and the mean scale factor evolves as:
\be
\rho\propto a^{-3(1+\frac{\qm}{6\qa})}, \quad a\propto t^{\frac{2}{3}(1-\frac{\qm}{6\qa})}.
\ee
Note that the $\afire$ is dynamically equivalent to the matter dominated epoch of \lcdm in the limit $\qm / \qa \rightarrow 0$ in which case $\Omega_m=1$, $\rho\propto a^{-3}$ and $a\propto t^{2/3}$.
In the same approximation as above the eigenvalues are:
\be
\left( -\frac{3}{2} + \frac{\qm}{4\qa}, \quad -1 +\frac{\qm}{2\qa}, \quad -\frac{3}{4} + \frac{\qm}{8\qa}, \quad -\frac{3}{4} + \frac{\qm}{8\qa}, \quad 3 + \frac{\qm}{2\qa} - \frac{\lambda}{2\qa}  \right).
\ee
The first four eigenvalues are negative under the assumptions of the approximation.  The last one is positive unless $\lambda/\qa \gtrsim 6$, i.e., $\lambda\gg 1$. Thus we must require $\lambda/\qa \lesssim 6$ to secure that $\vqmde$ is a saddle followed by an accelerated epoch.

\paragraph{Fixpoint $\afem$} This is a genuinely new type of scaling solution with matter, vector and quintessence. It exists in the intersection of $\Omega_A \ge 0$,  $\Omega_m \ge 0$ and $\Omega_V\ge 0$, which yields complicated algebraic expressions in terms of ($\qm,\qa,\lambda$). In principle it could provide interesting cosmological scenarios. Among the possibilities are accelerating scaling solutions, an ``almost de Sitter'' solution or alternatives to the matter dominated epoch. In appendix \ref{appendixa5} we have explored the cosmological viability of all these possibilities. It is clear from the expression for $\Sigma$ in table \ref{taban} that such alternatives would require fine-tuning in the parameters to secure a small shear ($|\Sigma|\ll1$) compatible with observations. In the appendix we show that even with an appropriate fine tuning, these alternatives are not cosmologically viable and can be ruled out. We conclude that (a5) can neither be responsible for the matter nor the dark energy dominated epoch.

\paragraph{Fixpoint $\vqde$} This is a scaling solution with vector and quintessence that can realize the late-time acceleration. It was first discovered in the context of inflation recently \cite{soda10,dimopoulos10}. Providing a counter-example to the cosmic no-hair theorem it has already generated much enthusiasm and attention. In the context of our model it plays the role as a possible future attractor together with $\qde$.  In section \ref{chdynamics} we shall see that in parameter regions where both $\qde$ and $\vqmde$ exists, the latter one will be dynamically selected. 

The existence of this fix-point relies on the following two conditions: $\Omega_A\ge0$ and $\Omega_V\ge0$. These two conditions can be written $\lambda^2+2\lambda \qa -4 \ge0$ and $\frac{4}{3}\le \frac{(2\qa+\lambda)^2}{\lambda^2+2\lambda\qa-2}$.  To avoid a large shear incompatible with observation we must require $|\Sigma|\ll1$ which is satisfied in the parameter region $\qa\gg\lambda$. For this case the existence condition simplifies to $2\lesssim \lambda\qa$.  Thus there is a large parameter region where (a6) exist and can be responsible for a small shear.

Under the assumption $\qa^2 \gg 1$ and $\qa\gg\lambda$  we expand the characteristics of the fix-point to first order in the small quantities $\frac{1}{\qa^2}$ and $\frac{\lambda}{\qa}$:
\be
\begin{split}
&X=-2\frac{1}{\qa}, \quad \Sigma=\frac{1}{3}\frac{\lambda}{\qa} - \frac{2}{3}\frac{1}{\qa^2}, \quad \Omega_A=\frac{1}{2}\frac{\lambda}{\qa} - \frac{1}{\qa^2} ,\quad \\ & \Omega_m=0, \quad \Omega_r=0, \quad \Omega_\phi =1- \frac{1}{2}\frac{\lambda}{\qa}+\frac{1}{\qa^2}, \quad \omega_\text{eff} = -1+ \frac{2\lambda}{3\qa} + 8 \frac{1}{\qa^2}.
\end{split}
\label{a6quantities}
\ee
Thus, to zeroth order, $\aseks$ looks like a de Sitter solution with $\Omega_\phi=1$ and $\weff=-1$. In section \ref{chCMB} we shall constraint the ratio $\lambda/\qa$ observationally by requiring a small shear compatible with observations.  Applying equations (\ref{eksakt1})-(\ref{eksakt3}) we find, in the same approximation, that the energy density and the mean scale factor evolves as:
\be
\rho\propto a^{-\frac{2\lambda}{\qa}}, \quad a\propto t^{\frac{\qa}{\lambda}}.
\ee
In the same approximation the eigenvalues read:
 \be
\left( -3+\frac{\lambda}{\qa}, \quad -4+2\frac{\lambda}{\qa},\quad -\frac{3}{2} + \frac{1}{2} \frac{\lambda}{\qa},\quad -\frac{3}{2} + \frac{1}{2} \frac{\lambda}{\qa}, \quad -3 + 2\frac{\lambda}{\qa}-2\frac{\qm}{\qa}   \right),
\ee
which shows that under the assumptions of the approximation, and the weak additional requirement that $|\qm|$ is not extremely large, $\vqde$ is a stable attractor.  In other words, $\vqde$ is the global attractor in the parameter region where it realizes a small shear compatible with observations.

\subsection{Parameter space \label{DSattractors}}
In this short section we shall use the results derived above to characterize parameter space in terms of the attractors. Figure \ref{figparameterspace} shows the parameter space for the slices $\qm=0.1$ and $\qm=-3.3$. Parameter space is divided into different regions each labeled by the fix-point that is an attractor in that region. Neglecting the boundaries we note that each point in parameter space has one and only one stable attractor. The green shaded region represents the part of parameter space where the attractor is accelerated, i.e., $\weff<-1/3$. Obviously, since we have ruled out the possibility of an accelerated saddle, the white region is ruled out since it cannot realize a dark energy dominated era.  In the next section we shall derive further, and stronger, bounds on the parameters by requiring consistency with the (observed) CMB quadrupole.  For $\qm=-0.1$ the plot is similar as the case $\qm=0.1$ in figure \ref{figparameterspace} except that (i6) is the stable fix-point also in the regions labeled (i5) and (a4).  Note that for the slice $\qm=-3.3$ the matter-quintessence scaling solution (i6) can realize an accelerated attractor. We will discuss this case further in section \ref{chpresent}.

Note that we have not considered the boundaries between the different regions of parameter space. For these special points a more careful analysis is needed (see, e.g., \cite{tiltedIII}). Furthermore, there may also be regions which contain other, more exotic, asymptotes, e.g., attracting closed curves. Although such possibilities in parameter space may pose mathematically interesting dynamical solutions we shall not attempt to identify them here as they merely represent special cases of our free parameters.\footnote{For more complicated Bianchi models such ``exotic'' behaviour is expected since such behaviours have been seen even in ordinary general relativity with perfect fluid \cite{tiltedVIIh,tiltedVIh}.}

\begin{figure}
\centering
\includegraphics[width=0.8\textwidth]{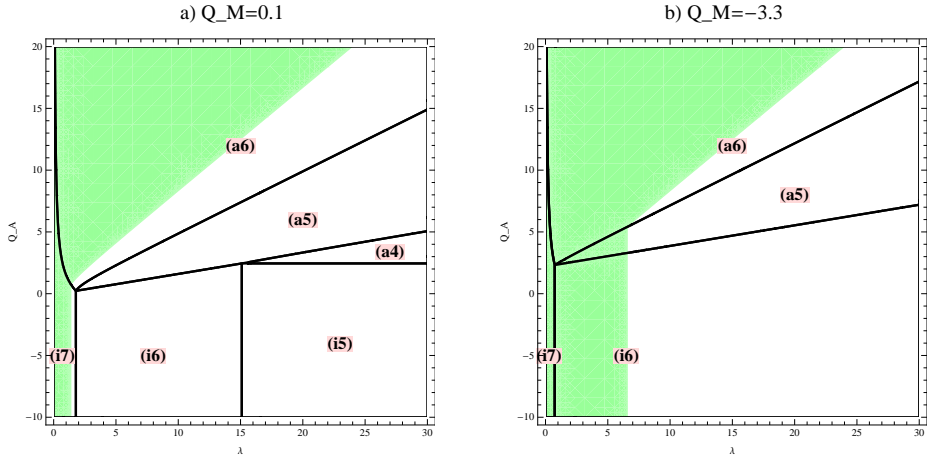}
\caption{Figure a) shows the parameter space ($\lambda$,$\qa$) of the model for the slice $\qm=0.1$. Each region is labeled by the fix-point that is an attractor in that region. The green shaded region is the accelerated part of the attractors.  For $\qm=-0.1$ the plot is similar except that (i6) is the stable fix-point also in the regions labeled (i5) and (a4). Figure b) shows the the same for the slice $\qm=-3.3$, a case to be discussed in section \ref{chpresent}.}
\label{figparameterspace}
\end{figure}

\subsection[Generality of the stability]{Generality of the stability of the inflationary attractors \label{chgeneric}}

In spite of the fairly complicated nature of the general system of equations, it is still possible to consider -- at least heuristically -- the stability of the exact solutions with respect to \emph{generic} inhomogeneous perturbations. In order to achieve this we will adopt the formalism for studying the dynamics of inhomogeneous
cosmologies ($G_0$ models) by employing expansion-normalised scale-invariant variables developed in \cite{UEWE}. This formalism uses orthonormal frames to write the Einstein field equations as an autonomous system of evolution equations and constraints. This formalism was successfully used to study the initial singuarity in \cite{UEWE}, and the instability of inhomogeneous perturbations in the more complicated Bianchi type VII$_h$ plane waves \cite{HCpp}. 

We should point out that although the equations in \cite{UEWE} are for a $\gamma$-law perfect fluid, some of the equations are purely geometrical in nature, i.e., stemming from equations like the Jacobi identity. The heuristic argument used here only depends on the background solutions and equations of  \emph{geometric origin}, thus the actual form of the matter source will be supressed. When that is said, the \emph{full} set equations for the scenario we are considering here is fairly straight-forward to derive by generalising the equations in \cite{UEWE} (see e.g., \cite{RU}).

The orthonormal frame $\{ {\bf e}_0,{\bf e}_{\alpha}\}$ can be expressed in local coordinates as 
\beq
{\bf e}_0=\frac{1}{N}(\dt-N^i\partial_i), \quad {\bf e}_{\alpha}=e_{\alpha}^{~i}\partial_i. 
\eeq
The Hubble-normalised frame is defined in general as follows: 
\beq
\dbold_0=\frac{1}{H}{\bf e}_0, \quad \dbold_{\alpha}=\frac{1}{H}{\bf e}_\alpha. 
\eeq
In addition to the deceleration parameter $q$, it is also necssary to introduce the spatial Hubble gradient $r_{\alpha}$:
\beq 
q+1\equiv-\frac{1}{H}\dbold_0H, \quad r_{\alpha} \equiv-\frac{1}{H}\dbold_\alpha H.
\eeq
Introducing the separable volume gauge: 
\beq
N=H^{-1}, \quad N^i=0, \quad \dot{U}_{\alpha}=r_{\alpha}, \label{temporalgauge}
\eeq
 enables us to write
\beq
\dbold_0=\dt, \quad \dbold_{\alpha}=E_{\alpha}^{~i}\partial_i,
\eeq
where $E_{\alpha}^{~i}\equiv e_{\alpha}^{~i}/H$.

The Hubble-normalised state vector for $G_0$ cosmologies  can then be given by 
\beq
{\bf X}=\left[E_{\alpha}^{~i},~r_{\alpha},~\Sigma_{\alpha\beta},~A^{\alpha},~N_{\alpha\beta},~\Omega,~ P,~\Pi_{\alpha\beta},~Q^{\alpha}\right]^T,
\eeq 
where  $\Sigma_{\alpha\beta}$ is the expansion-normalised shear, $A^{\alpha}$ and $N_{\alpha\beta}$ are the connection variables, and $\Omega$, $P$, $\Pi_{\alpha\beta}$, and $Q^{\alpha}$ are matter variables (the total expansion-normalised energy density, isotropic pressure, anisotropic stress and energy flux, respectively). The evolution equations and the constraints can now be written down in terms of ${\bf X}$ (see \cite{UEWE}). The equations can be written on the form
\beq
\dt {\bf X}&=& {\bf F}({\bf X},\partial_i{\bf X},\partial_i\partial_j{\bf X}),\label{eq:evol} \\
0&=& {\bf C}({\bf X},\partial_i{\bf X})\label{eq:constraints},
\eeq
where eqs.(\ref{eq:evol}) are the evolution equations and eqs. (\ref{eq:constraints}) are the constraints. 

The frame variables, $E_{\alpha}^{~i}$ decouple in the spatially homogeneous subset and thus have no direct consequence for the dynamics in this subset. However, for general inhomogeneous cosmologies they may have an impact on the dynamics. We are interested in perturbing  the fix-points found earlier, in particular, the stable fix-points (i7) and (a6) are of particular interest. Consider these fixed points, for which we can choose coordinates such that the frame variables are:
\[ E_{\alpha}^{~i}=\diag (E_1^x,~E_2^y,~E_3^z ).\]
The evolution equations and the commutator relations are 
 \beq
\dt E_{\alpha}^{~i}&=& \left(q\delta^{\beta}_{~\alpha}-\Sigma^{\beta}_{~\alpha}+\epsilon^{\beta}_{~\alpha\gamma}R^{\gamma}\right)E^{~i}_\beta \\
0&=& 2\left(\dbold_{[\alpha}-r_{[\alpha}-A_{[\alpha}\right)E_{\beta]}^{~i}-\varepsilon_{\alpha\beta\delta}N^{\delta\gamma}E_{\gamma}^{~i}, 
\eeq
In a Fermi-propagated frame, $R^\gamma=0$, and for the fix-points $q$ and $\Sigma_{\alpha\beta}=\diag(-2\Sigma,\Sigma,\Sigma)$ are constant; thus the evolution equations can be integrated to give:
\beq
E_1^x \propto e^{(q+2\Sigma)\tau}, \quad E_2^y \propto e^{(q-\Sigma)\tau} ,~E_3^z\propto e^{(q-\Sigma)\tau} .
\eeq

When perturbing the fix-point solutions with respect to the \emph{generic} inhomogeneous perturbations, the linearised equations will depend on the scale-invariant partial derivatives $\dbold_\alpha{\bf X}=E_\alpha^i\partial_i{\bf X}$, where $E_{\alpha}^i$ are the solutions above. In particular, if $f=f(\tau,x,y,z)$, then 
\beq
\dbold_1f\propto e^{(q+2\Sigma)\tau}\partial_xf,\qquad \dbold_2f\propto e^{(q-\Sigma)\tau}\partial_yf, \quad \dbold_3f\propto e^{(q-\Sigma)\tau}\partial_zf,
\eeq
to lowest order. 

To investigate the (in)stability of these spacetimes with respect to
general inhomogeneous perturbations we linearize the equations of
motion around the exact fix-point solutions (assuming that terms like $({\bf X}-{\bf X}_0)$, $\dbold_{\alpha}{\bf X}$, $\dbold_{\alpha}\dbold_{\beta}{\bf X}$, etc. are small). The interesting modes stem from the Hubble gradients $r_\alpha$, so to understand the dynamics of the perturbations we must linearise these equations explicitly. The evolution equation for $r_{\alpha}$ and the vorticity constraint read 
\beq
\dt r_{\alpha}&=& \left(q\delta^{\beta}_{~\alpha}-\Sigma^{\beta}_{~\alpha}+\varepsilon^{\beta}_{~\alpha\gamma}R^{\gamma}\right)r_\beta +\dbold_{\alpha}q, \\ 
0&=& \left[\varepsilon^{\alpha\beta\gamma}(\dbold_{\beta}-A_{\beta})-N^{\alpha\gamma}\right]r_{\gamma}.
\eeq
The deceleration parameter, $q$,  is given by
\[  q=2\Sigma^2+\frac 12(\Omega+3P)+\frac 23 A^{\alpha}r_{\alpha}-\frac 13\dbold_{\alpha}r^{\alpha}. \]
Lastly, we also need to consider the stability of the curvature variables (which can be interpreted as spatially homogeneous modes):
\beq
\dt A^\alpha&=& (q\delta^\alpha_{~\beta}-\Sigma^{\alpha}_{~\beta} +{\varepsilon^{\alpha}}_{\gamma\beta}R^\gamma)A^\beta+\frac 12\dbold_\beta(\Sigma^{\alpha\beta}+{\varepsilon^{\alpha\beta}}_\gamma R^\gamma), \\
\dt N^{\alpha\beta}&=& (q\delta^{(\alpha}_{~\delta}+2\Sigma^{(\alpha}_{~\delta}+2{\varepsilon_{\gamma\delta}}^{(\alpha}R^\gamma)N^{\beta)\delta}- \dbold_\gamma(\epsilon^{\gamma\delta(\alpha}\Sigma^{\beta)}_{~\delta}-{\delta^{\gamma(\alpha}}R^{\beta)} +\delta^{\alpha\beta}R^\gamma).
\eeq

Starting with the latter two equations, we note in the absence of inhomogeneous modes, that the \emph{fix-points considered here are stable to spatially homogeneous modes as long as}:
\beq
(q-4\Sigma)<0, \qquad (q+2\Sigma)<0.
\eeq
The inhomogeneous modes are a little bit more difficult to treat since there may  be solutions for which $\partial_i{\bf X}$ might diverge even if $E^i_\alpha \rightarrow 0$. We therefore need to make an assumption about the inhomogeneties. The inhomogeneous modes enter the evolution equations through $\dbold_{\alpha}{\bf X}=E^i_\alpha\partial_i{\bf X}$. Let us compare with inhomogeneous modes $f_{\bf k}=\exp(i{\bf k}\cdot {\bf x})$, where ${\bf k}=(k_x,k_y,k_z)$ are in general time dependent:
\[ \dbold_{\alpha}f_{\bf k}=i(E^i_\alpha k_i)f_{\bf k}.\]
Whether such modes grow or decay depends on the term $E^i_\alpha k_i$. If 
\[ (q+2\Sigma)<0, \quad (q-\Sigma)<0,\]
then the $E^i_{\alpha}$ will decay and, if $k_i$ does not grow too quickly, then $E^i_\alpha k_i\rightarrow 0$. Moreover, are these conditions satisfied, one can see from the equations above that as $\tau\rightarrow \infty$:
\[ (E^i_{~\alpha},r_\alpha,\dbold_\alpha X)\rightarrow (0,0,0),\]
(in line with \cite{UEWE}, this is the condition for  \emph{future asymptotic silence}). 
In particular, $k_i$ growing can be interpreted as the inhomogeneities "collapsing" (in terms of the coordinates). However, they only have cosmological significance if $(E^i_\alpha k_i)$ do not remain small. This implies that inhomogeneties which have modes $k_i$ not growing too rapidly will experience a "homogeneisation", or smoothing out, during an inflationary phase for which  
$(q+2\Sigma)<0$ and $(q-\Sigma)<0$. 

For our most interesting solutions, the fix-points (i7) and (a6), we have $(q,\Sigma)\approx(-1,0)$ and $(q,\Sigma)\approx (-1,10^{-5})$, and thus all of the bounds are fulfilled. Consequently, these are stable against homogeneous pertubations, as well as inhomogeneous perturbations provided that the modes $k_i$ do not grow too quickly. 

We should also point out that the equations used here are purely geometric (aside from the actual expression for $q$), thus the heuristic perturbation argument is model independent and fairly robust. Thus, we expect similar stability for other models with similar behaviour. 

\section{Cosmology of the strong vector coupling regime \label{chstrongvector}}
Based on the dynamical system analysis above we shall now identify a cosmologically very interesting region of parameter space. This is the \emph{strong vector coupling regime} defined by 
\be
|\qa|\gg 1, \quad |\qa|\gg |\qm|, \quad |\qa| \gg \lambda. 
\label{strongvector}
\ee
The goal of this section is to understand the cosmology of this parameter region and put bounds on the parameters of the model. As we shall see, the strong vector coupling regime yields viable cosmologies with important differences compared to Standard Coupled Quintessence (SCQ). Given appropriate initial conditions in the radiation dominated era ($\weff\simeq 1/3$),  a subsequent matter dominated epoch ($\weff\simeq 0$) will be realized before entering the accelerated era ($\weff < -1/3$).  In general the dynamically selected solutions in the matter and dark energy dominated epochs will have a small expansion anisotropy. The associated shear variable is automatically small in the strong vector coupling regime. Consistency with the CMB quadrupole will provide upper bounds on the fractions $|\qm/\qa|$ and $|\lambda/\qa|$. No additional fine-tuning in any of the parameters is needed to avoid a dominating and observationally unacceptable shear.

As we shall see, the strong vector coupling regime is characterized by only a handful of fix-points. These are $\rde$, $\vqmde$, $\qmde$, $\iseks$, $\vqde$ and $\isyv$ which we analyzed individually in the previous section.  In this section we shall analyze the cosmological trajectories that connect these solutions.  We shall see that the cosmology can easily be understood qualitatively in terms of these fix-points. 

In section \ref{chinitialconditions} we identify the initial conditions yielding viable cosmologies. In \ref{chdynamics} we study the possible trajectories starting with viable initial conditions. In \ref{chpresent} we consider a special case where the present universe, represented by $\iseks$, is the global attractor. In \ref{chCMB} we consider the CMB quadrupole and put bounds on the parameters of the model.  Finally, in \ref{chsound}, we consider the sound horizon.  

\subsection{Initial conditions \label{chinitialconditions}}
Among the fix-point solutions in our model several have $\weff = 1/3$ and are therefore natural candidates for the radiation dominated era.  These are the ``concordance'' radiation saddle $\rde$, which has $\Omega_r=1$, and the scaling solutions $\itre,\,\ifire,\, \ato$, and $\atre$. In this section we shall argue that the scaling solutions can be ruled out, and that cosmologically viable initial conditions must be represented by the concordance saddle $\rde$.  We show this by assuming that the matter dominated epoch is realized by either $\afire$ or $\ifem$.\footnote{In section \ref{chphasespace} we ruled out $\iseks$ and $\afem$ as candidates for the matter dominated epoch.} 

Viable initial conditions must provide a matter dominated era squeezed in between the radiation dominated era and the accelerated epoch. The matter dominated epoch must last several e-folds such that matter perturbations can grow and eventually collapse non-linearly on scales up to approximately $1$ Mpc. Obviously, this puts strong bounds on the initial conditions. First note that the matter dominated fix-points $\vqmde$ and $\qmde$ have $\omega_\phi=1$, i.e., the kinetic part of $\phi$ is responsible for the scalar field energy.  From (\ref{seqV}) it follows that the potential energy of the scalar field will scale approximately as $\Omega_V \propto a^3$ during a matter dominated era (assuming $\lambda \lesssim 1$). Thus, in order to have a matter dominated era lasting several e-folds, $\Omega_V$ must be extremely small at the time of matter-radiation equivalence.  This observation implies that a viable candidate for the radiation dominated era must have $\omega_\phi =1$.  Otherwise, i.e., if $\omega_\phi < 1$, the $\Omega_V$ ``handed over'' from the radiation dominated era will be way too large (for any reasonable  parameters $\lambda$,$\qm$,$\qa$) to provide a matter dominated epoch sufficiently stable to last from $z\sim 3000$ until $z\sim 1$. We can therefore rule out $\ifire$ and $\ato$ as candidates for the radiation dominated era. 

Next, let us consider the scaling solution $\itre$ which has $\omega_\phi=1$ and $\Omega_\phi = 1/(6\qm^2)$. The bound on $\Omega_\phi$ at big bang nucleosynthesis is \cite{bean01}:
\be
\Omega_\phi^\text{bbn} < 0.045,
\label{bbn}
\ee
which means that $\qm^2 > 3.7$ given that (i3) is responsible for the radiation era.  In standard coupled quintessence such a large $|\qm|$ can be ruled out since it is impossible to realize the subsequent matter dominated era via $\qmde$ which only exists for $\qm^2<3/2$. Apparently, our more general model can realize a subsequent matter dominated era via $\afire$ with a large $|\qm|$. It turns out, however, to be impossible is to get a sufficiently stable radiation dominated era that lasts, as a minimum, from big bang nucleosynthesis all the way to redshift $z\sim 3000$ (where radiation-matter equivalence is supposed to occur). To see this, first note that $|\qa| \gg |\qm|$ is required to have a small shear ($\Sigma \sim \qm / \qa$) during $\afire$. Furthermore $\qa$ and $\qm$ must have the same sign for $\afire$ to exist, i.e., $\qa/ \qm \gg 1$. In section \ref{chCMB} we shall see that consistency with the CMB quadrupole requires $\qa/\qm \gtrsim 10^5$. Next, consider the field equation for $\Omega_A$, equation (\ref{seqA}), that includes a term $-\qa X$ on the right hand side.  Evaluated at (i3) this term is $\qa / \qm$. Such a large term would cause the vector to grow absurdly fast, $\Omega_A \propto a^{10^5}$, at least initially when it is subdominant.  After the initial rapid grow, simulations shows that $\Omega_A$ will start to oscillate and in fact stay subdominant. This will, however, trigger $\Omega_m$ to start growing and the radiation era is replaced by the matter era within a few e-folds.   To have a radiation era stable from BBN to matter-radiation equality would require an unreasonable fine tuning in the initial condition for $\Omega_A$. The only reasonable possibility is to put $\Omega_A=0$ as initial condition, but in that case $\afire$ cannot be realized and consequently there is no matter dominated epoch. Clearly, we can rule out (i3) as a candidate for the radiation dominated era.

Finally, let us consider the scaling solution $\atre$. From the conditions for existence given  in section \ref{DSisotropic}, it follows that the fix-point does not exist in the strong vector coupling regime defined in (\ref{strongvector}). Even outside the strong vector coupling regime (a3) can be ruled out since the existence of $\ifem$ is incompatible with the BBN bound (\ref{bbn}) and the shear of $\afire$ is large in the parameter region where both (a3) and (a4) exists. 

Using only weak assumptions, we have showed that initial conditions must be chosen close to the concordance radiation saddle $\rde$ to get a viable cosmology.  It follows that the potential energy of the scalar field will scale approximately as a cosmological constant during the radiation era ($\Omega_V\propto a^{4}$), and in fact also during the matter era ($\Omega_V\propto a^{3}$). It follows that $\Omega_V$ must be extremely small initially.  Exactly the same sort of fine-tuning is required also in other quintessence models with the exponential-type potential.

\subsection{Dynamics \label{chdynamics}}
Although our model is phenomenologically very rich, we shall now see that in the strong vector coupling regime the dynamics is characterized by only a handful of fix-points and can more easily be understood qualitatively.  In addition to the constraints on the parameters coming from the strong vector coupling regime, see (\ref{strongvector}), we shall assume that $|\qm|$ and $\lambda$ are smaller than unity in this section.  Phase space is then characterized by the fix-point solutions $\rde$, $\vqmde$, $\qmde$, $\vqde$ and $\qde$. Properties of these fix-points are summarized in table \ref{case1}.  

\begin{table}
\centering
\footnotesize
\begin{tabular}{l*{5}{l}} \\
\hline\hline
           & $\rde$ & $\qmde$ & $\vqmde$ & $\qde$  & $\vqde$    \\ \hline
Existence         & - & $\qm^2 \le \frac{3}{2}$ & $4\qm\qa \gtrsim 1$ & $\lambda^2 \le 6$ & $2\lesssim \lambda\qa $ \\
Stability            & saddle  & saddle & saddle & attractor or & attractor \\
& & & & saddle & \\ 
$\weff$     &    $1/3$ & $\frac{2\qm^2}{3}$ & $\frac{\qm}{6\qa}$ & $-1+\frac{\lambda^2}{3}$ & $-1+ \frac{2\lambda}{3\qa} + 8 \frac{1}{\qa^2}$   \\
$\Sigma$ &   $0$ & $0$  & $\frac{2\qm}{3\qa}-\frac{1}{6\qa^2} $ & $0$ & $ \frac{1}{3}\frac{\lambda}{\qa} - \frac{2}{3}\frac{1}{\qa^2}$  \\    
$\rho$      & $\propto a^{-4}$ & $\propto a^{-3-2\qm^2}$ &   $\propto a^{-3(1+\frac{\qm}{6\qa})}$ & $\propto a^{-\lambda^2}$ & $\propto a^{-\frac{2\lambda}{\qa}}$    \\
$a$           & $\propto t^{1/2}$ & $\propto t^{2/(3+2\qm^2)}$ &    $\propto t^{\frac{2}{3} (1-\frac{\qm}{\qa})}$ & $\propto t^{2/\lambda^2}$ & $\propto t^{\frac{\qa}{\lambda}}$   \\
\hline\hline
\end{tabular}
\caption{The fix-points characterizing the strong vector coupling regime. Several quantities are approximated using the assumptions of the strong vector coupling regime (\ref{strongvector}).}
\label{case1}
\end{table}

We should emphasize that although we consider $\lambda$ and $|\qm|$ smaller than unity in this section, there are also large parameter regions yielding viable cosmologies when they are larger than unity. For $\lambda$ and/or $|\qm|$ larger than unity, the situation is a bit more complicated, however, since one must carefully avoid the problematic regions of parameter space discussed in section \ref{chphasespace}, e.g., regions where $\qmde$, $\afem$ or $\iseks$ represent matter dominated attractors (none of these undesirable situations exist when $|\qm|$ and $\lambda$ are smaller than unity). Another difference is that $\ifem$ and $\isyv$ do not exist when $\qm$ and $\lambda$ are large (see table \ref{case1}). An interesting possibility, to be discussed in section \ref{chpresent}, that occurs when $|\qm|$ and $\lambda$ are larger than unity, is that $\iseks$ represents the present universe.  Apart from special cases involving (i6), the following discussion covers the main features of the cosmology of the strong vector coupling regime. Several of the scenarios discussed in this section also exist when $|\qm|$ or $\lambda$ are larger than unity.

Viable solutions must realize a matter dominated era squeezed in between the radiation and dark energy dominated epochs;  in other words the following sequence must be realized:
\be
(\weff \simeq 1/3) \quad \rightarrow \quad (\weff \simeq 0) \quad \rightarrow \quad (\weff < -1/3).
\ee
In our model this sequence can be achieved in various ways:   
\begin{equation} 
 \Big( \rde \Big) \quad \rightarrow \quad 
\left(
\begin{array}{ccc} \vqmde  \\  \text{or} \\ \qmde 
\end{array} 
\right) 
\quad \rightarrow \quad
\left(
\begin{array}{ccc} \vqde  \\  \text{or} \\ \qde 
\end{array} 
\right). 
\end{equation}
As shown schematically, the matter dominated epoch can be realized either by the anisotropic scaling solution $\vqmde$ or the isotropic scaling solution $\qmde$, while the dark energy dominated epoch can be realized either by the anisotropic scaling solution $\vqde$ or the isotropic solution $\qde$. This yields four qualitatively different trajectories. Examples of each of these trajectories are given in figure \ref{bigsimulation}.  In general the anisotropic solutions are dynamically preferred over the isotropic ones. If both $\vqmde$ and $\qmde$ exists, the former will be dynamically selected in the matter dominated epoch. Similarly, if both $\vqde$ and $\qde$ exists, the former will be dynamically selected in the dark energy dominated epoch.\footnote{There may be a transient period close to $\qde$ before $\vqde$ is reached. An example of this is shown in simulation b) of figure \ref{bigsimulation}.}  We can therefore easily determine the trajectories from the existence conditions for $\vqmde$ and $\vqde$.\footnote{The isotropic fix-points are guaranteed to exists since we consider the case that both $|\qm|$ and $\lambda$ are smaller than unity.}  In the strong coupling regime the existence conditions are very simple to lowest order; $\vqmde$ exists when $\qm$ and $\qa$ have similar signs, while $\vqde$ exists when $\qa>0$.  Thus each of the four trajectories occupies a region of parameter space of roughly equally large size. We shall comment on each of them later. First let us observe a common feature of all trajectories, namely that the vector field scales similarly as radiation, i.e., $\Omega_A\propto\Omega_r=\text{constant}$, during the radiation dominated epoch. This is due to the fact that the coupling term $\qa X$ on the right hand side of (\ref{seqA}) is suppressed during $\rde$. Hence the vector is effectively uncoupled and therefore scales as an ordinary Maxwell field, $\rho_A \propto a^{-4}$.\footnote{Here, we have assumed that the shear is negligible during the radiation dominated epoch.} The fact that the vector is stable during the radiation dominated epoch suggests that inflation is a natural candidate for the initial conditions for the vector. Then, let us comment on each of the four types of trajectories.

\begin{figure}[h]
\centering
\includegraphics[width=0.94\textwidth]{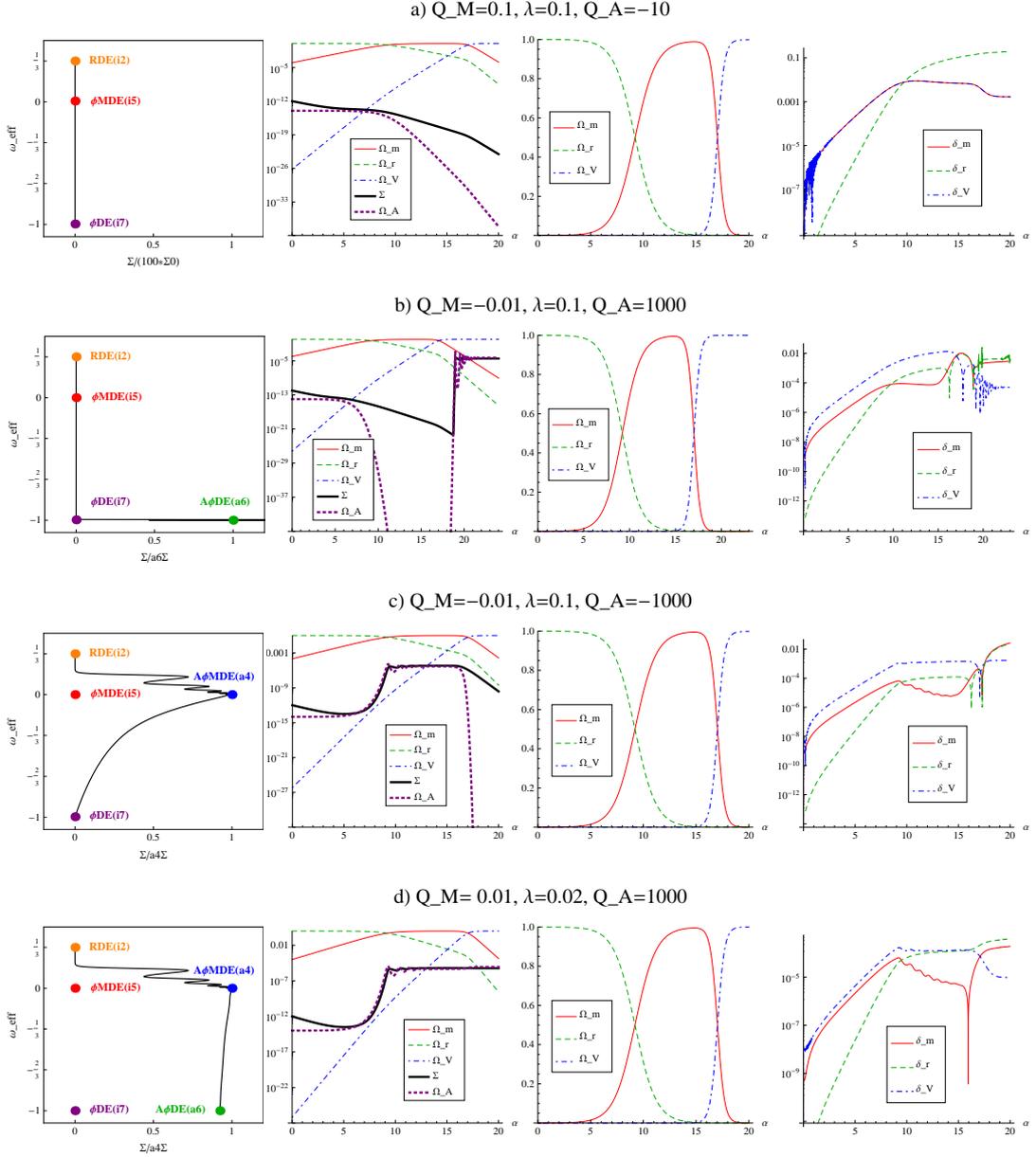}
\caption{The phase flow and dynamics for four different sets of parameters (written in the figure), but with the same initial conditions close to the concordance radiation saddle $\rde$. The four set of parameters give examples on each of the four different types of trajectories described in section \ref{chdynamics}. In each simulation (labeled a-d) the first figure in the corresponding column shows the phase flow (displaying the relevant fix-points that exists there), the second figure is a logarithmic plot of the time evolution of the dynamical variables, the third figure is similar with linear y-axis (it is then no point to plot the vector and shear since they are too small visually), and the fourth figure compares $\Omega_m$, $\Omega_r$ and $\Omega_V$ to a \lcdm simulation that starts with the same initial conditions ($\delta_i = |\Omega_i / \Omega_i^{\lambda\text{CDM}} -1|$ ).  In the phase flow diagram the shear axis is normalized such that the fix-point that has the largest shear is positioned at $x=1$ (for instance a4$\Sigma$ is the shear at $\vqmde$ which can be looked up in table \ref{case1}).}   
\label{bigsimulation}
\end{figure}

\paragraph{$\rde\rightarrow\qmde\rightarrow\qde$} This trajectory is close to isotropy and characterize the standard coupled quintessence model \cite{amendola}. See simulation a) of figure \ref{bigsimulation} for an example of this trajectory. It is realized if neither $\vqmde$ nor $\vqde$ exist, i.e., in the parameter region where $1 \gtrsim 4\qm\qa $ and $2\gtrsim \lambda\qa $.   As explained above, the energy density of the vector field scales similarly as radiation during $\rde$.  During $\qmde$ and $\qde$ the vector typically decays rapidly, $\Omega_A\propto a^{-|1-4\qa\qm|}$ and $\Omega_A\propto a^{-|4-2\lambda\qa|}$ respectively. Since this type of trajectory exists in SCQ, the phenomenology at the perturbative level is well understood \cite{amendola}.

\paragraph{$\rde\rightarrow\qmde\rightarrow\vqde$} This trajectory is close to isotropy in both the radiation and matter dominated epochs, but anisotropic in the dark energy dominated epoch. See simulation b) of figure \ref{bigsimulation} for an example of this trajectory. It is realized if $\vqmde$ does not exist and $\vqde$ exists, i.e., in the parameter region where $1 \gtrsim 4\qm\qa $ and $2\lesssim \lambda\qa $. During $\qmde$ the vector scales as $\Omega_A\sim a^{-1+4\qa\qm}$ and therefore decays rapidly for typical parameters consistent with the above condition. The trajectory will therefore typically have a transient period around the isotropic fix-point $\qde$ (since some time is needed for $\Omega_A$ to grow) before settling at the anisotropic attractor $\vqde$. The trajectory is therefore usually more precisely described by $\rde\rightarrow\qmde\rightarrow\qde\rightarrow\vqde$.  Notice that this is the case for simulation  b) of figure \ref{bigsimulation} where there is an isotropic period in the beginning of the dark energy dominated epoch. This period is short or entirely vanishing in the special case where $4\qm\qa$ is just slightly smaller than unity.  In general, however, the cosmology of this trajectory is indistinguishable from SCQ up to today ($\Omega_\phi\simeq 0.7$); the universe will first become significantly anisotropic in the future.

\paragraph{$\rde\rightarrow\vqmde\rightarrow\qde$} This trajectory describes a universe which is close to isotropy during the radiation dominated epoch, anisotropic during the matter dominated epoch and close to isotropy again in the dark energy dominated epoch.  In other words, anisotropization occurs during the transition to the matter dominated epoch, while isotropization occurs under the transition to the accelerated era. To the best of our knowledge, this is genuinely new behavior not predicted by any other known model.  See simulation c) of figure \ref{bigsimulation} for an example of this. Such trajectories are realized when $\vqmde$ exists and $\vqde$ does not exist, i.e., in the parameter region where $1 \lesssim 4\qm\qa $ and $2\gtrsim \lambda\qa $.

\paragraph{$\rde\rightarrow\vqmde\rightarrow\vqde$} This trajectory describes a universe which is anisotropic both in the matter and dark energy dominated epochs. Anisotropization occurs during the transition to the matter dominated epoch, while the transition to the dark energy dominated epoch represents a transition between two anisotropic scaling solutions. Also this is genuinely new behavior.  Such trajectories are realized when both $\vqmde$ and $\vqde$ exists, i.e., in the parameter region where $1 \lesssim 4\qm\qa $ and $2\lesssim \lambda\qa $.  For an example of this trajectory, see simulation d) of figure \ref{bigsimulation}. In that simulation the shear has approximately the same value during both $\vqmde$ and $\vqde$. In general however, depending on the model parameters, the shear may be very different during the matter and dark energy dominated epochs.

\subsection{The present universe as a global attractor \label{chpresent}}
In the previous sections we have demonstrated that, given some conditions on the parameters and the initial condition, our model is, in many ways, dynamically very similar to the concordance model.  We shall now see that also a more exotic, yet interesting, scenario can be realized. A global accelerated attractor with $\Omega_\phi\simeq 0.7$ and $\Omega_m \simeq 0.3$ is often discussed as an interesting possibility in the literature \cite{hebecker00, amendola01}. In this scenario the present universe has already reached the global attractor and we do not live in the middle of a transition between two cosmic epochs.  It is sometimes argued that such a scenario reduces the coincidence problem of the standard model, namely the question why we happen to live at a very special moment in the cosmic history \cite{amendola01}.  

The fix-point $\iseks$ is a global attractor with $\Omega_\phi \simeq 0.7$ and $\Omega_m \simeq 0.3$ when $(\qm,\,\lambda)=(-3.3,\, 2.2)$ and $\qa<0$, see case b) of figure \ref{figparameterspace}.  In standard coupled quintessence it is impossible to realize a matter dominated epoch with these parameters since $\ifem$ only exists for $\qm^2<3/2$ \cite{amendola}. In the frame work of that model one therefore needs to introduce a field dependent matter coupling, $\qm(t)$, which is small during $\ifem$, but eventually grows large enough to realize the desired attractor \cite{amendola01}. In our model, however, there is no need to introduce time variation in $\qm$ since the matter dominated epoch can be realized by $\afire$ even with a large $|\qm|$. Note that $\afire$ exists and has a small shear with $\qm=-3.3$ given the condition $\qa \ll \qm$.  In figure \ref{specialattractor} we show how our model sucssessfully realizes the considered scenario through the sequence $\ito \rightarrow \afire \rightarrow \iseks$.  As far as we know there is no other model with a simple stationary coupling that can realize the present universe as a global attractor.

Although our model can realize the present universe as a global attractor without any problems on the background level, it is expected that the large matter coupling will lead to problems for the perturbed universe.  When the matter dominated epoch is realized by $\ifem$ there is a strong bound $|\qm|<0.13$ which comes from a full CMB analysis \cite{amendola03}. As we shall see in section \ref{chsound}, it is expected that the bounds on $|\qm|$ is somewhat relaxed when the matter dominated era is realized by $\afire$. Still we believe $|\qm|=3.3$ is way too large, but a full study of perturbations is required to verify this.  

\begin{figure}[h]
\centering
\includegraphics[width=1.0\textwidth]{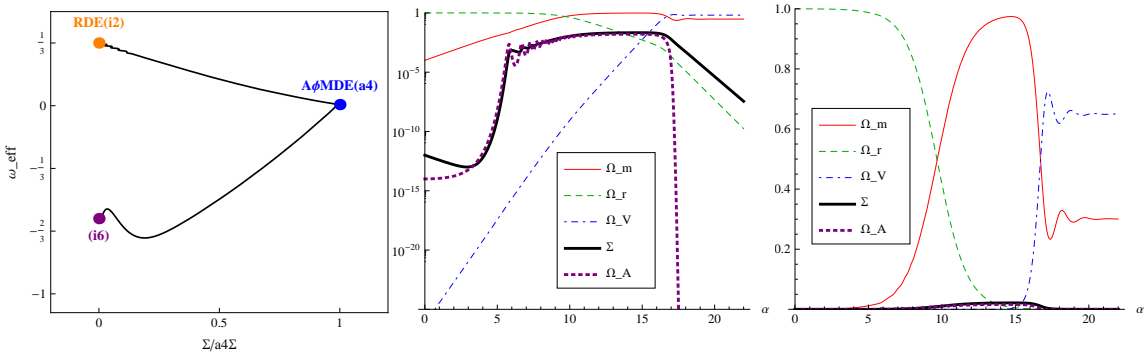}
\caption{Simulation where the present universe is a  a global attractor represented by $\iseks$ with $\Omega_\phi\simeq 0.7$ and $\Omega_m\simeq 0.3$.  The parameters used are $\qm=-3.3$, $\lambda=2.2$ and $\qa=-100$. }
\label{specialattractor}
\end{figure}

\subsection{Bounds from CMB quadrupole \label{chCMB}}
In this section we shall constrain the model parameters observationally using the CMB. Calculating the expectation values of the CMB anisotropies in the Bianchi I spacetime is in general a comprehensive task. Perturbation theory of the Bianchi I spacetime is to a large extent worked out (see \cite{pereira07,gumrukcuoglu07,koivisto08a,nakamura11} and references therein), but a complete numerical implementation in a Boltzmann code is so far not achieved. When the shear is very small, however, one can treat the spacetime as a perturbation about the FLRW metric and consider only the leading corrections to the CMB. In this approach (used for instance in \cite{barrow97b,jaffe05,campanelli06,campanelli07,koivisto08a}) one consistently accounts for the anisotropic redshift of the photons emitted at the last scattering surface, but neglects any correction beyond this zero order effect. Note that the redshift in Bianchi models depends on the direction of propagation since the spacetime background is anisotropic.  The total temperature anisotropy field $\Delta T/\left<T\right>$ therefore consists of one part coming from inflation $\Delta T^i/\left<T\right>$, which is similar as in the FLRW metric, and a ``redshift part'' $\Delta T^a/\left<T\right>$ due to the anisotropic metric, i.e., $\Delta T/\left<T\right> = \Delta T^i/\left<T\right> + \Delta T^a/\left<T\right>$.  In appendix \ref{appcmb}  we show that the leading term in $\Delta T^a/\left<T\right>$ is a quadrupole term $\mathcal{Q}^a$ which we shall refer to as the \emph{shear quadrupole}. The quadrupole of $\Delta T^i/\left<T\right>$ we shall denote $\mathcal{Q}^i$ and refer to as the \emph{inflationary quadrupole}. All shear multipoles higher than the quadrupole are strongly suppressed and can be neglected whenever the shear is sufficiently small to be compatible with observations.  

The total (observed) quadrupole $\mathcal{Q}$ depends on $\mathcal{Q}^i$, $\mathcal{Q}^a$ and their mutual orientation. Depending on the orientation the total quadrupole may be larger or, more interestingly, smaller than the one coming from inflation. The 7-year WMAP data gives the value \cite{jarosik10}:
\be
\mathcal{Q}^2 \simeq 200 \left( \frac{\mu K}{T_\text{cmb}} \right)^2 \sim \left(10^{-5} \right)^2,
\ee 
while the expectation value for the $\Lambda$CDM concordance model is:
\be
\left< (\mathcal{Q}^i)^2 \right> \simeq 6 \mathcal{Q}^2.
\ee
Even taking into account the cosmic variance, $\sqrt{2/5} \left<(\mathcal{Q}^i)^2 \right>$, the observed value is suprisingly low. It is well-known that for a shear quadrupole roughly of the same size as the inflationary quadrupole, the total quadrupole can be lowered to match the observed value \cite{campanelli06,campanelli07}. This requires a suitable orientation of $\mathcal{Q}^i$ relative to $\mathcal{Q}^a$ which, in fact, is not improbable, see \cite{campanelli11}. 

 Let us use this to constrain the parameters of our model. Although the possibility $\mathcal{Q}^a\sim \mathcal{Q}^i$ is attractive, since it can explain the low observed quadrupole, it is clear that we must also allow  parameters predicting $\mathcal{Q}^a \ll \mathcal{Q}^i$. In the latter case the predicted total quadrupole is essentially equal to the one predicted by $\Lambda$CDM and the low observed value must be accepted as a statistical fluke. It is also clear that $\mathcal{Q}^a$ cannot be several orders of magnitude larger than $\mathcal{Q}^i$ since this would give, for any orientation, a large total quadrupole incompatible with observations. We shall use that the shear quadrupole must be around the same order of magnitude as the inflationary quadrupole or smaller: 
\be
\mathcal{Q}^a \lesssim 10^{-5}.
\label{quadrupolebound}
\ee

Then, let us relate the shear quadrupole to the parameters of our model.  For convenience we fix the coordinates such that the scale factors are unity today, i.e., $a_\parallel(t_0)=a_\perp(t_0)=1$ and therefore $\alpha(t_0)=\sigma(t_0)=0$.  The shear quadrupole is then given by (see appendix \ref{appcmb} for a derivation): 
\be
\mathcal{Q}^a= \frac{12}{5\sqrt{3}} \left| \sigma_{\text{dc}}  \right|,
\label{Q2}
\ee
where $\sigma_{\text{dc}}=\sigma(t_{\text{dc}})$ is the metric function defined in (\ref{metric}) evaluated at the decoupling time. Note that $\mathcal{Q}^a$ is denoted $Q^a_2$ in appendix \ref{appcmb}.  To put bounds on the model parameters we shall derive an analytical expression for $\mathcal{Q}^a$. To derive a robust order of magnitude estimate we assume that the shear is given by fix-point $\afire$ all the way from the decoupling time ($z\sim 1100$) up to today.\footnote{Since the transition to dark energy dominance must have happened quite recently, around $z\simeq 0.3$, this is a quite good approximation although we know that the system (verified by simulations) will be closer to $\aseks$ at $z=0$.} Since $\Sigma$ is constant (as in any fix-point), it follows from definition (\ref{shear}) that $\sigma_\text{dc} = \Sigma \alpha_\text{dc}$.\footnote{Note that, since $|\alpha_\text{dc}| \simeq \log 1100$ is of order unity, $|\Sigma|\ll1$ implies $|\sigma_\text{dc}| \ll 1$.}  Inserting into (\ref{Q2}), using the value for the shear in (\ref{a4quantities}), we get 
\be
\mathcal{Q}^a =  \frac{2|\alpha_\text{dc}|}{5\sqrt{3}} \frac{(4\qm\qa -1)}{\qa^2},
\label{Qa2}
\ee
where the scale is related to the redshift by $\alpha_\text{dc} = -\log (z_\text{dc}+1)$. The constraint (\ref{quadrupolebound}) then leads to: 
\be
\frac{4\qm\qa -1}{\qa^2} \lesssim 10^{-5},
\label{bounda4lang}
\ee
where we have omitted prefactors. Note that this bound holds only if $4\qm\qa \in (1,\infty)$ as required by the existence of  $\afire$. For typical parameters consistent with the existence of (a4) the constraint (\ref{bounda4lang}) can therefore be rewritten as: 
\be
 \boxed{\left| \qm/\qa \right|  \lesssim 10^{-5}}.
\label{bounda4}
\ee
For special parameters such that $4\qm\qa$ is close to unity one should use (\ref{bounda4lang}); otherwise (\ref{bounda4}) holds. 

So far we have neglected the shear during the accelerated era, $\Sigma_\text{(a6)}$. Let $t_*$ denote the time of equivalence between matter and dark energy, i.e., $\Omega_\phi(t_*)=\Omega_m(t_*)\simeq0.5$.  Since $t_*$ corresponds to a low redshift, $z_*\simeq 0.3$, $\Sigma_\text{(a4)}$ will provide the main contribution to the shear quadrupole if $\Sigma_\text{(a4)} \simeq \Sigma_\text{(a6)}$. If the matter era is realized by $\qmde$, however, like simulation b) of figure \ref{bigsimulation}, or if $\Sigma_\text{(a6)} \gg \Sigma_\text{(a4)}$, then it is $\Sigma_\text{(a6)}$ that will provide the main contribution to $\mathcal{Q}^a$. It is clear that we will get further bounds on the model parameters by considering the shear of $\vqde$. Using the value for the shear in (\ref{a6quantities}) as an estimate for the shear between $t_*$ and $t_0$ we get the following estimate for the shear quadrupole: 
\be
\mathcal{Q}^a =  \frac{4|\alpha_*|}{5\sqrt{3}} \frac{\lambda \qa - 2}{\qa^2}.
\label{Qa2a6}
\ee
The constraint (\ref{quadrupolebound}) then leads to:
\be
\frac{\lambda \qa - 2}{\qa^2} \lesssim 10^{-4},
\label{bounda6lang}
\ee
This bound holds only if $\lambda\qa \in (2,\infty)$ as required by  the existence of (a6).  For typical parameters consistent with the existence of (a6) the constraint (\ref{bounda4lang}) can be rewritten as: 
\be
\boxed{ \left| \lambda/\qa \right|  \lesssim 10^{-4}}.
\label{bounda6}
\ee
For special parameters such that $\lambda\qa/2$ is close to unity one should use (\ref{bounda6lang}); otherwise (\ref{bounda6}) holds. Note that the bound on $|\lambda/\qa|$ is relaxed with an order of magnitude compared to $|\qm/\qa|$.

\subsection{Sound horizon \label{chsound}}
In the previous section we showed that consistency with the observed CMB quadrupole requires $|\qm|/|\qa|\lesssim 10^{-5}$ if $\afire$ is responsible for the matter era. The smallness of $|\qm|/|\qa|$ have several consequences.  Firstly it implies that the expansion is almost shear-free.  Secondly it implies  that the background  expansion history, which goes as $a\propto t^{(2/3)(1- \qm/6\qa)}$ during $\afire$, is practically indistinguishable from the standard \lcdm matter era where $a\propto t^{2/3}$. This is in contrast to $\ifem$ where $a\propto t^{2/(3+2\qm^2)}$ and the bound $\qm\lesssim 0.13$ \cite{amendola03} allows for a significantly different expansion history. 

The sound horizon at decoupling time is governed by the background expansion history.  Assuming that the matter era is realized by a fix-point that includes the kinetic part of the scalar field, one can derive the following approximation for the sound horizon at decoupling  \cite{amendola,amendola99,DEboka}:
\be
r_s \simeq r_0 z_\text{dc}^{\frac{1}{2}\qm X},
\label{sound}
\ee 
where $r_0$ is the standard \lcdm sound horizon, $z_\text{dc}$ is the redshift at decoupling time and $X$ is the dimensionless variable for the kinetic part of the scalar field defined in (\ref{X}) (which must be evaluated in the given fix-point). For $\ifem$ one gets $r_s = r_0 z_\text{dc}^{-\qm^2}$. For $\qm=0.1$, for instance, this gives a sound horizon $7\%$ smaller than in \lcdmm. Since both photons and baryons are uncoupled in our model, we can calculate the sound horizon in exactly the same way. In particular, the exact background solution corresponding to \emph{any} matter fix-point in our model, i.e., (\ref{eksakt1})-(\ref{eksakt3}) with $\beta_i=\beta_m$, leads to the same expression for the sound horizon (\ref{sound}).   For $\afire$ (\ref{sound}) gives $r_s = r_0 z_\text{dc}^{-\qm/4\qa}$ which means that the sound horizon is indistinguishable from the standard horizon. In the concordance model we know that the multipole location of the first acoustic peak is roughly inversely proportional to the sound horizon, $l_\text{peak}\propto r_s^{-1}$. Conjecturing that the same relation also holds in our model, it follows that $\afire$ cannot move the position of the first acoustic peak by means of a modified sound horizon. This is in contrast to $\ifem$ which can move $l_\text{max}$ several percents.  

Since it is the fraction $\qm/\qa$ that determines the background expansion history in $\afire$, the magnitude of $\qm$ itself is not dynamically relevant. It is therefore expected that the bounds on $|\qm|$ separately is somewhat relaxed compared to the bound $|\qm|<0.13$ in standard coupled quintessence.  A full analysis of perturbations is required to confirm this.

\section{Summary and discussion \label{chconclusion}}
In this work we have explored the cosmological consequences of a quintessence field with couplings to a Maxwell-type vector field as well as to dark matter.  Both couplings have separately been studied extensively  in the literature, but have until now not been considered simultaneously.  In a dark energy context, motivated by higher-dimensional theories such as string theories, it is natural to consider the case where both types of couplings are present. We have showed that genuinely new behaviors arise due to the double coupling, and we have identified an interesting parameter region, corresponding to a strong vector coupling regime, yielding exciting and viable cosmologies close to the \lcdm limit. Below we shall summarize our main results and discuss possible extensions and generalizations of the model.  

Doubly Coupled Quintessence (DCQ) can be viewed as a generalization of Standard Coupled Quintessence (SCQ) \cite{amendola}.  This is reflected in the dynamical system where all the \emph{isotropic} fix-points also exist in the simpler SCQ model. The dynamical system is characterized by 13 physically different fix-points, 7 isotropic and 6 anisotropic (see tables \ref{tabis} and \ref{taban} respectively).   The properties of these fix-points are controlled by three dimensionless parameters $\lambda$, $\qa$ and $\qm$  which determine the shape of the scalar potential and the strength of the couplings to the vector and matter fields, respectively.  We identified a large parameter region, the strong vector coupling regime (see (\ref{strongvector})), which yields viable cosmologies. In this region, the dynamics is characterized by only a handful of fix-points and can easily be understood qualitatively. In section \ref{chinitialconditions} we established that viable cosmologies must have a radiation dominated epoch close to the isotropic concordance radiation saddle $\rde$. This is required to have a sufficiently stable matter dominated epoch squeezed in between the radiation and dark energy dominated epochs.  

Interestingly, both the matter and the dark energy dominated epochs can be realized by anisotropic solutions, i.e., scaling solutions where a subdominant, but stable, vector field sources a small anisotropy in the expansion rate. $\vqmde$ is a genuinely new matter dominated scaling solution with a small stable vector field and a small quintessence field which can be responsible for the matter dominated epoch.  $\vqde$ is a quintessence dominated scaling solution with a small stable vector field. This fix-point was first discovered in the context of inflation \cite{soda10,dimopoulos10}, but in the framework of our model it plays the role as the future attractor responsible for the late-time acceleration. There are also isotropic alternatives for the matter and dark energy dominated epochs, provided by $\qmde$ and $\qde$ respectively.  These isotropic fix-points characterize the dynamics of SCQ. Thus we have anisotropic and isotropic alternatives both in the matter and the dark energy dominated epochs. Which trajectory that is dynamically preferred depends on the parameters in the theory, and we used the dynamical system approach to address this. In a certain parameter region where there are both anisotropic and isotropic alternatives for a given cosmic epoch, the anisotropic solution will be dynamically selected. It follows that the isotropic solutions are realized only in parameter regions where the anisotropic solution do not exist. When $|\qm|$ and $\lambda$ are smaller than unity there are four different types of trajectories which characterize the dynamics of the strong vector coupling regime: 
\begin{itemize}
\item isotropic both in the matter and the dark energy dominated epochs (equivalent to SCQ): $\qmde\rightarrow \qde$,
\item isotropic matter dominated epoch and anisotropic dark energy dominated epoch: $\qmde\rightarrow \vqde$,
\item anisotropic matter dominated epoch and isotropic dark energy dominated epoch (isotropization occurs under transition to the dark energy dominated epoch): \newline $\vqmde\rightarrow \qde$,
\item anisotropic in both the matter and dark energy dominated epochs: $\vqmde\rightarrow \vqde$.
\end{itemize}
All trajectories occupy roughly equally large regions of parameter space. Simulations for all types of trajectories are shown in figure \ref{bigsimulation}, while further details are discussed in section \ref{chdynamics}.  The rich dynamical behavior clearly distinguishes the model from other proposed late-time anisotropic models.  In models where the anisotropic expansion is sourced by an isotropy-violating dark energy field \cite{koivisto05,beltran07,koivisto07,koivisto08a,koivisto08,akarsu08,battye09,campanelli11}, the anisotropization typically occurs during the transition to the dark energy dominated epoch ($z\lesssim 0.3$). This behavior is a reminiscent of the second special case $\qmde\rightarrow \vqde$ of our model. Another possibility often considered in the literature is isotropy violation by a cosmic magnetic field  (usually assumed to have a primordial origin) \cite{barrow97, barrow97b, berera03, campanelli06, campanelli07, demianski07, campanelli09,kahniashvili08}.  The energy density of such a magnetic field scales similarly as radiation ($\propto a^{-4}$) which means that the dimensionless density parameter will be stable during the radiation dominated epoch while it will decay ($\propto a^{-1}$) during the matter dominated epoch.  Thus the anisotropy will die out during the matter dominated epoch and will be negligible at low redshifts. Although there are some similarities, this is clearly different from the third special case $\vqmde \rightarrow \qmde$ of our model where the anisotropy during the matter dominated epoch \emph{arise} during the transition from the radiation dominated epoch and then is stable all the way to the transition to the dark energy dominated epoch where it dies out. In the fourth type of trajectory the transition to the dark energy dominated epoch represents a transition between two anisotropic scaling solutions, also a new type of behavior.  

Let us also briefly comment on a fifth type of trajectory that occurs if we allow $|\qm|$ and $\lambda$ to be larger than unity. For $(\qm,\,\lambda)=(-3.3,\, 2.2)$ the isotropic fix-point $\iseks$ represents a global attractor with $\Omega_\phi\simeq 0.7$ and  $\Omega_m \simeq 0.3$. Thus we have the exotic possibility that the present state of the universe is the global attractor. This scenario is not possible to realize in the framework of SCQ since it is impossible to realize a matter dominated epoch. In our doubly coupled model, however, the matter dominated epoch will be realized by $\vqmde$. It is therefore possible to realize a cosmologically viable sequence with the present universe as the global attractor. Nevertheless, problems are expected at the perturbative level, see the discussion in section \ref{chpresent}.

The CMB quadrupole is very sensitive to shear and we used this to constrain the parameters of the model. More concretely we put upper bounds on the ratio of the strength of the matter coupling to the strength of the vector coupling, and on the ratio of the parameter in the quintessence potential and the strength of the  vector coupling. As derived in \ref{chCMB} the upper bound is $|\qm/\qa|\lesssim 10^{-5}$ if the matter dominated epoch is realized by $\vqmde$, and $|\lambda/\qa|\lesssim 10^{-4}$ if the dark energy dominated epoch is realized by $\vqde$.  The surprisingly low quadrupole observed in WMAP is in fact not improbable if $|\qm/\qa|$ or $|\lambda/\qa|$ lays around the  upper bounds such that the shear quadrupole is of the same order of magnitude as the inflationary quadrupole.  Given a suitable orientation (which is not improbable), the observed quadrupole will then be lower than the inflationary quadrupole in harmony with observations \cite{campanelli06,campanelli07,campanelli11}. 

Our model is dynamically equivalent to \lcdm in the limit $\qm / \qa \rightarrow 0$, $\lambda/ \qa \rightarrow 0$. In this limit the anisotropic solutions $\vqmde$ and $\vqde$ become equivalent to the matter and dark energy dominated epochs of \lcdmm, respectively. Therefore, although DCQ is very different from \lcdm qualitatively,  the strong bounds on the model parameters imply that the quantitative differences on the background level are extremely small.  For example the mean scale factor evolves as $a \propto t^{(2/3) (1-\qm/\qa)}$ during $\vqmde$ which is practically indistinguishable from \lcdm where $a\propto t^{2/3}$. As verified in the simulations in figure \ref{bigsimulation},  there are only minor differences from \lcdm on the background level for an epoch realized by an anisotropic solution (when the constraints on $|\qm/\qa|$ and $|\lambda/\qa|$ are satisfied).  A full study of perturbations is required for a more complete understanding of the phenomenological consequences of the anisotropic epochs. For $\qmde$ the phenomenology at the perturbative level has been explored in the literature and the consequences for structure formation and the CMB are well understood \cite{amendola}.  The upper bound on the matter coupling is $|\qm|<0.13$ given that $\qmde$ is responsible for the matter dominated epoch \cite{amendola03}. Interestingly, we expect that the bound on $|\qm|$ is somewhat relaxed compared to this if the matter dominated epoch is instead realized by $\vqmde$ since in that case the sound horizon at decoupling time is indistinguishable from \lcdm (see section \ref{chsound}). 

There are several natural ways to generalize the model.  For instance it would be interesting to consider time variations in the coupling constants and/or other types of scalar potentials that exhibit tracking behavior \cite{steinhardt99,amendola03}. It would also be interesting to see if the model can be (minimally) modified to realize a non-standard radiation dominated epoch. Since the constraint on a primordial shear is much weaker than at late times \cite{campanelli11b}, it is of particular interest to check if such a minor modification could allow for a radiation dominated era realized by one of the anisotropic scaling solutions $\ato$ and $\atre$. To better understand the observational signatures of DCQ a full analysis of the perturbations is necessary. This work will be carried out in the near future. 

Finally, it should again be emphasized that the Maxwell-type vector field considered in this paper represents a possible dark component of the universe and must not be confused with the photons of the standard model of particle physics.  Note that if our considered field was identified with photons, the coupling to quintessence would correspond to an extremely rapid time variation in the fine structure constant, $\alpha_\text{fs} \propto f(\phi)^{-2}$, clearly incompatible with observational constraints (the upper bound on time variation of the fine structure constant between decoupling time and today is at the one-percent level \cite{uzan02,rocha03}). This rules out the possibility that our considered vector field could be identified with a possible homogenous component of large-scale cosmic magnetic fields. 

To conclude, we have showed that the model is dynamically very rich, and can realize various scenarios that violate isotropy while being close enough to the \lcdm limit to be taken seriously. An important  lesson learned is that although the universe \emph{looks} like \lcdm based on present observations, it certainly does not mean that the universe \emph{is} \lcdmm. Cosmological observations have forced us to introduce two new dark fields beyond the standard model of particle physics. The present energy density of both dark energy and dark matter exceed the energy density of baryons by approximately one order of magnitude. We should therefore not be surprised if future observations lead to discoveries of other less dominant types of fields with too little energy or too weak couplings to be seen in present data.

\acknowledgments{We thank Tomi S. Koivisto, Sigurd K. N{\ae}ss, Thiago S. Pereira, Federico R. Urban, Hans A. Winther and Kei Yamamoto for valuable discussions. DFM thanks the Research Council of Norway FRINAT grant 197251/V30. DFM is also partially supported by project CERN/FP/123615/2011 and PTDC/FIS/111725/2009.}

\newpage

\appendix

\section{Eigenvalues \label{cheigen}}
The stability analysis in section \ref{chphasespace} is based on the following eigenvalues of the matrix of linear perturbations around the fix-points: 

\begin{align}
&\text{Fixpoint (i1$\pm)$}:\quad (2,\; 0,\; 2 \mp 2 \sqrt{6} \qa,\; 3 \pm \sqrt{6} \qm,\; 6 \pm \sqrt{6} \lambda). \label{eigeni1} \\
&\text{Fixpoint (i2):} \;\qquad  (4,\; -1,\; -1,\; 1,\; 0). \label{eigeni2} \\
&\text{Fixpoint (i3):}\;\quad  \left(-1,\; 2 \frac{\qa}{\qm},\; -\frac{1}{2} - \frac{\sqrt{2 - 3 \qm^2}}{2 \qm},\; -\frac{1}{2} + 
  \frac{\sqrt{2 - 3 \qm^2}}{2 \qm},\; 4 - \frac{\lambda}{\qm} \right). \label{eigeni3} \\
&\text{Fixpoint (i4):} \quad  \left( -1, \; 8 \frac{\qa}{\lambda},\; 1 - \frac{4\qm}{\lambda},\; - \frac{1}{2} - \frac{\sqrt{
 64 - 15 \lambda^2}}{2 \lambda},\; -\frac{1}{2} + \frac{\sqrt{
 64 - 15 \lambda^2}}{2 \lambda}
 \right). \label{eigeni4} \\
& \text{Fixpoint (i5):} \notag \\
& \Bigg( -\frac{3}{2} + \qm^2, \; -\frac{3}{2} + \qm^2,\; -1 + 2 \qm^2,\; -1 + 2 \qm (2 \qa + \qm),3 +  2 \qm (\qm - \lambda) \Bigg).  \label{eigeni5} \\
& \text{Fixpoint (i6):} \notag \\
& \Bigg( -\frac{3}{2} + \frac{3\qm}{2(\lambda - \qm)}, \; -1+\frac{3 \qm}{\lambda - \qm} , \; -1+\frac{3 (2 \qa+\qm)}{\lambda - \qm} ,\notag \\  & \qquad\qquad\quad -\frac{3(\lambda-2\qm)}{4(\lambda-\qm)}\left[ 1 \pm \sqrt{1+\frac{8(3+\lambda(\qm-\lambda))(3+2\qm(\qm-\lambda))}{3(\lambda-2\qm)^2}} \right]
 \Bigg). \label{eigeni6} \\
& \text{Fixpoint (i7):} \notag \\
& \left( -3 - \lambda \qm  + \lambda^2, \; -3 +\frac{\lambda^2}{2}, \; -3 +\frac{\lambda^2}{2} , \; -4 + \lambda^2, \; -4 + 2 \lambda \qa  + \lambda^2 \right). \label{eigeni7} \\
&\text{Fixpoint (a1$\pm)$}: \left( 0, \quad 2, \quad 2 \mp 4 \sqrt{1 - \frac{1}{6} X^2} - 2 \qa X, \quad 3 + \qm X, \quad 6 +  \lambda X \right). \label{eigena1} \\
&\text{Fixpoint (a2):} \qquad \left( 1-\frac{4 \qm}{\lambda }, \quad   -\frac{1}{2}\pm\frac{1}{2\lambda}\sqrt{A\pm BC}  \right), \label{eigena2} \\
&\text{where:} \notag \\
& A = \left(32+48 \qa^2-8 \qa \left(2+3 \qa^2\right) \lambda -\left(7+12 \qa^2\right) \lambda ^2\right), \notag \\
& B = -4 \sqrt{\left(2+3 \qa^2\right)  }, \notag \\
& C =  \sqrt{12 \qa^4 \lambda ^2-8 \qa \lambda  \left(-4+\lambda ^2\right)+2 \left(-4+\lambda ^2\right)^2+12 \qa^3 \lambda  \left(4+\lambda ^2\right)+3 \qa^2 \left(16+\lambda ^4\right)}. \notag  
\end{align}

\newpage

\begin{align}
& \text{Fixpoint (a3):} \qquad  \left(4-\frac{\lambda }{\qm}, \quad -\frac{1}{2} \pm \frac{1}{4\qm}\sqrt{A\pm B}  \right), \label{eigena3} \\
&\text{where:} \notag \\
& A =   4+6 \qa^2-8 \qa \left(2+3 \qa^2\right) \qm-4 \left(1+3 \qa^2\right) \qm^2, \notag \\
& B =   2\sqrt{ \left(2+3 \qa^2 \right) \left(2+3 \qa^2+8 \qa \left(2+3 \qa^2\right) \qm+4\qm^2 \left(-2+9 \qa^2 +12 \qa^4\right) + C \right)}, \notag  \\
& C =  16 \qa \left(-2+3 \qa^2\right) \qm^3+4 \left(2+3 \qa^2\right) \qm^4. \notag \\
& \text{Fixpoint (a4):} \notag \\
& \Bigg( 3 - 3A \left(6+9 \qa^2+7 \qa \qm\right), \quad 8 - 6 A \left(6+9 \qa^2+7 \qa \qm\right), \notag \\
& \qquad \frac{3}{4} \left(-1 + \qm (\qa+3 \qm)A \pm ABC  \right), \label{eigena4} \\ 
& \qquad 3A \left(4+6 \qa^2+6 \qa \qm+4 \qm^2-\qa \lambda -3 \qm \lambda \right)  \Bigg), \notag \\
&\text{where:} \notag \\
& A = 1/ \left(4+(2 \qa+\qm) (3 \qa+\qm) \right), \notag \\
& B = -2 \sqrt{\left(-2-3 \qa^2-2 \qa \qm+\qm^2\right)},  \notag \\ 
& C = \sqrt{ \left(-8+32 \qa^3 \qm+13 \qm^2+2 \qa \qm \left(9+4 \qm^2\right)+\qa^2 \left(-11+32 \qm^2\right)\right)}.  \notag \\
& \text{Fixpoint (a5): extremely lengthy algebraic expressions in the parameters } (\qm,\qa,\lambda).   \notag \\
& \text{Fixpoint (a6):} \notag \\
& \Bigg( -3A (8+(6 \qa-\lambda ) (2 \qa+\lambda )), \quad -8A \left(4+6 \qa^2+\qa \lambda -\lambda ^2\right), \notag \\
& \qquad  -\frac{3}{2}+ 3A \lambda  (2 \qa+\lambda )8+(2 \qa+\lambda ) (6 \qa+\lambda ) \pm AB, \label{eigena6} \\ 
& \qquad  -3A (8+(6 \qa+4 \qm-3 \lambda ) (2 \qa+\lambda )) \Bigg), \notag \\
&\text{where:} \notag \\
& A = 1/ \left(8+(2 \qa+\lambda ) (6 \qa+\lambda )\right), \notag \\
& B =-\frac{3}{2} \sqrt{\left(-8-12 \qa^2-4 \qa \lambda +\lambda ^2\right) (-72+(2 \qa+\lambda ) (17 \lambda +2 \qa (-19+4 \lambda  (2 \qa+\lambda ))))}. \notag
\end{align}

\section{Non-viability of fix-point (a5) \label{appendixa5}}
In section \ref{DSanisotropic} we stated that (a5) cannot be responsible for neither the matter dominated nor the accelerated epoch.  Here we show why.

First we consider the case $\lambda \gg \qm$ in which case $\omega_\text{eff} \simeq 0$. Let us check if this matter dominated scenario is cosmologically viable. In the limit $\lambda \gg \qm$ the shear is $\Sigma\simeq -\frac{1}{4}+\frac{3}{2}\frac{\qa}{\lambda}$.  Since the eigenvalues of the Jacobian matrix are extremely complicated, it is convenient to neglect higher order terms in the small quantities $a = \frac{\qm}{\lambda}$, $b=-\frac{1}{4} + \frac{3}{2} \frac{\qa}{\lambda}$. Note that $a\ll 1$ is required to have $\omega_\text{eff}\simeq 0$, while $b\ll 1$ is required to have a small shear $|\Sigma|\ll1$. To first order in $a$ and $b$ the conditions for existence are:
\begin{align}
a+b\ge 0, \label{a5_ex1} \\
-1+\frac{1}{3}\lambda^2 - \frac{1}{3} \lambda^2 ( 2a+b) \ge 0, \label{a5_ex2} \\
2-\lambda^2(a-\frac{1}{3}b) \ge 0 \label{a5_ex3}.
\end{align}
When $\lambda$ is of order unity, this simplifies to $\lambda^2\gtrsim 3$ and $a+b\ge 0$.  For any $\lambda^2 \gtrsim 3$ it is possible to satisfy (\ref{a5_ex1}-\ref{a5_ex3}) simultaneously for small $a$ and $b$. Thus we have showed that it is possible for (a5) to represent a matter dominated solution with a small shear. To say whether this possibility is cosmologically viable we need to know the stability.  In our considered approximation the real part of the eigenvalues are:
\begin{align}
\left(-1+3a, \quad  -\frac{3}{2} + \frac{11}{2} a + 4 b, \quad -4(a+b), \quad -\frac{3}{4}(1-a), \quad -\frac{3}{4}(1-a)  \right)
\end{align}
Here we have assumed $\lambda^2>\frac{24}{7}$ to simplify the eigenvalues (several terms then become imaginary). The existence conditions allows for a slightly smaller value, $\lambda^2 \gtrsim 3$, but from $\Omega_m = 1-\frac{3}{\lambda^2} + \mathcal{O}(a) + \mathcal{O}(b)$ one see that $\lambda^2 \gg 3$ is needed to have matter as the dominant component.\footnote{For $\lambda^2\simeq 3$ the dominating component is the scalar field with equal kinetic and potential part such that the total equation of state is zero. }  Notice that the third eigenvalue is negative when (\ref{a5_ex1}) is satisfied. The other eigenvalues are also negative in our considered approximation. It follows that the matter dominated solution with a small shear is a stable attractor. The solution can therefore not escape from the matter era into the accelerated era. Thus (a5) does \emph{not} provide a cosmologically viable alternative to the matter dominated era.

Next we check if the fix-point can drive acceleration. Acceleration is realized if 
\be
\omega_\text{eff} = -\frac{\qm}{\qm-\lambda} \in [-1,-1/3).
\ee
Since we consistently assume $\lambda>0$ in this paper, we have acceleration if and only if  $\qm<-\frac{1}{2}\lambda$.\footnote{Note that for $\qm>\lambda$ we have $\omega_\text{eff}<-1$ and the fix-point do not exist.}  It is possible to satisfy the conditions for existence together with the condition for acceleration only if $\qm<-\sqrt{2}/2$.  Thus a small matter coupling, $|\qm| \ll 1$ is incompatible with (a5) driving acceleration.  For the case $\qm<-\sqrt{2}/2$ there is a restricted region of parameter space that is  compatible with the conditions for existence and acceleration. As we shall now see, however, this region does not provide a viable cosmology.   

First we check the interesting special case of an ``almost de Sitter'' solution, i.e., $\omega_\text{eff}\simeq -1$ and $|\Sigma| \ll 1$. It is convenient to use the three quantities $\qm$, $a=-1-\frac{3}{2}\frac{\qa}{\qm}$ and $b=\frac{\lambda}{\qm}$.  We assume that the latter two are small ($a\ll1$, $b\ll1$) which is equivalent to $\Sigma \ll 1$ and $\omega_\text{eff}\simeq -1$.  From the constraint $\Omega_m > 0$ it follows that $-2-\qm^2(\frac{4}{3}a +b) \ge 0$ to first order in $a$ and $b$. Thus the solution requires a large matter coupling, $\qm^2 \gg 1$. This rules out the possibility for (a5) to represent an ``almost'' deSitter solution.

Next we consider the possibility of an accelerated scaling solution with $\Omega_\phi \neq 0$ and $\Omega_m\neq 0$.  In this case it is convenient to use the three quantities $\qm$, $\omega_\text{eff}$ and $\Sigma$. Recall that we consistently consider the case $\lambda>0$. In terms of our considered quantities this translates into $\qm<0$. Let us assume acceleration and a small shear, i.e., $\omega_\text{eff}\in (-\frac{1}{3},-1]$ and $\Sigma\ll1$.  Given these assumptions the constraint $\Omega_V \ge 0$ is automatically satisfied while the conditions $\Omega_A > 0$ and $\Omega_m>0$ are equivalent to $\Sigma\ge 0$ and $\qm^2\ge 3 \frac{\omega_\text{eff}^2}{1+\omega_\text{eff}}$. The latter condition implies a minimum value on $|\qm|$.  Tuning the model to the concordance model by setting $\omega_\text{eff} = -0.7$ implies $\qm^2 > 4.9$. In order to have a small shear we must require:
\be
\Sigma = \frac{\lambda-6\qa-4\qm}{4(\qm-\lambda)} \ll 1.
\label{sheara5}
\ee
Using $\weff=-0.7$ it follows from (\ref{sheara5}) that $\qm/\qa\simeq-42/31$ which is incompatible with a matter dominated epoch via $\vqmde$. Also note that the condition $\qm^2>4.9$ derived above is incompatible with a matter dominated epoch via $\qmde$ (the existence condition for $\qmde$ is $\qm^2\le3/2$ ). Thus it is impossible to realize a matter dominated epoch in the parameter region where (a5) represents an accelerated epoch.  This rules out (a5) as a candidate for the matter dominated era.

\section{CMB quadrupole in the Bianchi type I spacetime \label{appcmb}}
Here we shall derive the leading effects of the shear on the CMB. Other derivations can be found in \cite{barrow85,koivisto08a}. Let us consider the most general Bianchi type I spacetime:
\be
ds^2=-dt^2+a(t)^2dx + b(t)^2 dy^2 + c(t)^2dz^2.
\ee
Our first goal is to derive a formula for the redshift of photons coming from the last scattering surface.  Since the expansion is anisotropic the redshift will be a function of the direction of the incoming photon:
\be
z(\hat{\mathbf p}) = \frac{\lambda_0}{\lambda_{dc}},
\ee
where $\lambda_{dc}$ and $\lambda_0$ is the wavelength of photons at decoupling (last scattering surface) and today, respectively. We will derive this formula using an analogy to photons propagating in the flat FLRW metric. First consider the geodesic equation:
\be
\frac{du^{\mu}}{d\lambda} = -\Gamma^\mu_{\alpha\beta} u^\alpha u^\beta,
\ee
where the only non-vanishing Christoffel coefficients are
\be
\Gamma^0_{ii} = a_i \dot a_i, \quad \Gamma^i_{0i}=\Gamma^i_{i0} = \frac{\dot a_i}{a_i}.
\ee 
Note that if the tangent vector of a photon is parallel to a coordinate line initially, the photon will continue to propagate along that coordinate line. For example if the photon's four-velocity initially is parallel to the x-axis, $u^2=u^3=0$, then $\frac{du^2}{d\lambda}=\frac{du^3}{d\lambda}=0$ according to the geodesic equation. We shall now use this observation to find an expression for the redshift as a function of the direction of the received photon.  Let the photon's direction of propagation be defined by the unit (three) vector
\be
\hat p^i=(\hat p_x,\hat p_y,\hat p_z)=(\sin\theta \cos\phi,\; \sin\theta \sin\phi,\; \cos\theta).
\ee
We now introduce new coordinates $\widetilde x^\mu=(t,\tilde x,\tilde y,\tilde z)$ which are (spatially) rotated relative to $x^\mu$ such that the $\tilde x$ - axis is parallel to the photon's direction of propagation:
\be
\tilde x = \hat p_x x + \hat p_y y + \hat p_z z.
\label{ctr}
\ee
We let $\tilde a$, $\tilde b$ and $\tilde c$ be the scale factors in the new coordinate chart. The coordinate transformation defined by (\ref{ctr}) determines the metric component $\tilde g^{11}$:
\be
\tilde a^{-2} = \displaystyle \sum_i \frac{\hat p_i^2}{a_i^2}.
\label{sf}
\ee 
According to the discussion above the photon will propagate along the $\tilde x$ coordinate line. The photon's propagation is therefore subject to the null-condition:
\be
ds^2 = -dt^2 + \tilde a^2 d\tilde x^2.
\ee
This is mathematically equivalent to a photon propagating in the flat FLRW metric. Thus the redshift is
\be
1+z(\hat{\mathbf p}) = \frac{\tilde a_0}{\tilde a},  
\label{rs}
\ee 
where $\tilde a$ is the scale factor at the time of emission and $\tilde a_0$ today. For convenience we use $a_0=b_0=c_0=1$ which implies $\tilde a_0=1$. From (\ref{sf}) and (\ref{rs}) we then get an expression for the redshift as a function of the direction of propagation for the incoming photon emitted at the decoupling time:
\be
1+z(\hat{\mathbf p}) = \sqrt{\displaystyle \sum_i \frac{\hat p_i^2}{(a_i)^2_\text{dc}}},
\ee
where the subscript indicates that the scale factors are evaluated at the decoupling time.  This is our desired formula, the redshift parametrized by the direction of the photon. From here we follow more or less the derivation in \cite{koivisto08a}.  It is convenient to rewrite the redshift: 
\be
1+z(\hat{\mathbf{p}}) = \frac{1}{a} \sqrt{1+\hat{p}_y^2 \epsilon + \hat{p}_z^2 \eta},
\ee
where 
\be
\epsilon = \left( \frac{a}{b}\right)^2 -1, \quad  \eta = \left( \frac{a}{c}\right)^2 -1.
\ee
Note that $|\epsilon|$ and $|\eta|$ essentially are the ``eccentricities'' squared. For a small shear the eccentricities will be small and we can expand to first order in $\epsilon$ and $\eta$. First we calculate the mean redshift over all directions to first order in $\epsilon$ and $\eta$: 
\be
\left< z(\hat{\mathbf{p}}) \right> = \frac{1}{4\pi} \int d\Omega_{\hat{\mathbf p}} z(\hat{\mathbf{p}}) = -1+\frac{1}{\overline a} + \mathcal{O}(\epsilon^2,\eta^2),
\label{averagez}
\ee
where $\overline a = (abc)^{1/3}$ is the geometric mean of the scale factors. In the main text we defined an isotropic redshift $z$, see (\ref{isoredshift}). Notice that to first order in the perturbations $\left< z(\hat{\mathbf{p}}) \right>=z$. Thus our defined isotropic redshift has an interpretation as the averaged redshift when the shear is small. 

The homogenous background temperature field will therefore be anisotropic:
\be
\begin{split}
T^a(\hat{\mathbf p}) &= \frac{T_{dc}}{1+z(\hat{\mathbf{p}})} \\
&= a T_{dc} (1-\frac{1}{2}\hat{p}_y^2 \epsilon - \frac{1}{2} \hat{p}_z^2 \eta) + \mathcal{O}(\epsilon^2,\eta^2),
\end{split}
\ee
where $T_{dc}$ is the decoupling temperature.  In the following we shall omit the notation $\mathcal{O}(\epsilon^2,\eta^2)$, but  it is clear that we will neglect all second order terms. The total temperature field is $T=T^a + \Delta T^i$ where $\Delta T^i$ is the (inhomogeneous) part coming from inflation. The average (total) temperature is:
\be
\left<T\right> = \frac{1}{4\pi} \int d\Omega_{\hat{\mathbf p}} T^a(\hat{\mathbf{p}}) = aT_\text{dc} \left(  1-\frac{1}{6}(\epsilon+\eta)   \right).
\ee
The shear temperature anisotropy field is therefore:
\be
\delta(\hat{\mathbf p}) \equiv \frac{\Delta T^a(\hat{\mathbf p})}{\left< T \right>}=  1 - \frac{T^a(\hat{\mathbf p})}{\left< T \right>} = \frac{1}{2}\hat{p}_y^2 \epsilon + \frac{1}{2} \hat{p}_z^2 \eta - \frac{1}{6}(\epsilon + \eta), 
\ee
or in terms of the angular variables:
\be
\delta(\theta,\phi)= \frac{1}{2} \epsilon \sin^2\!\theta \sin^2\!\phi  + \frac{1}{2} \eta \cos^2\!\theta - \frac{1}{6}(\epsilon + \eta).
\ee
We shall now expand the field in the spherical harmonics:
\be
\delta(\theta,\phi) = \displaystyle\sum_{l} \displaystyle\sum_{m=-l}^{l} a_{lm} Y_{lm} (\theta,\phi),
\ee
where, due to the orthogonality of the spherical harmonics, the coefficients are given by:
\be
\begin{split}
a_{lm} &= \int d\Omega_{\hat{\mathbf p}} \delta(\hat{\mathbf p})  Y^*_{lm}.
\end{split}
\ee
The multipoles are defined
\be
Q^a_l = \sqrt{\frac{1}{2\pi} \frac{l(l+1)}{(2l+1)}  \displaystyle\sum_{m=-l}^{l}|a_{lm}|^2}.
\ee
To first order in $\epsilon$ and $\eta$ there are only non-vanishing coefficients for $l=2$:
\be
a_{20} = \frac{1}{3}\sqrt{\frac{\pi}{5}}(2\eta-\epsilon), \quad a_{21}=a_{2-1}=0, \quad a_{22}=a_{2-2}=-\sqrt{\frac{\pi}{30}}\epsilon,
\ee
which gives
\be
Q^a_2 = \frac{2}{5\sqrt{3}}\sqrt{\epsilon^2+\eta^2-\epsilon \eta}.
\ee
Thus for our axisymmetric spacetime (\ref{metric}) the quadrupole is\footnote{The normalization of the scalefactors to unity today implies that $\sigma(t_0)=0$.}
\be
Q^a_2 = \frac{12}{5\sqrt{3}} \left| \sigma_{\text{dc}}  \right|.
\ee
To second order in our small quantities we also find a hexadecapole:
\be
Q^a_4 =  \frac{36}{28} \sqrt{\frac{31}{15}} \sigma_\text{dc}^2.
\ee
Note that $Q^a_4 \sim (Q^a_2)^2$. More generally, for even multipoles we have $Q^a_{2n} \sim (Q^a_2)^{n}$ while the odd multipoles vanish exactly $Q^a_{2n+1} =0$.  Thus the leading correction on the CMB anisotropy field is a quadrupole coming from the shear that will mix with the inflationary quadrupole $Q^i_2$. Since observationally we know that $Q_2 \lesssim 10^{-5}$ it is clear that we can neglect the correction to the hexadecapole and any higher multipoles.

\bibliographystyle{JHEP}
\bibliography{refs}

\end{document}